\documentclass[11pt]{article}

\usepackage{epsfig}
\usepackage{subfig}
\usepackage{graphicx}
\usepackage{epstopdf}
\usepackage{amsfonts}
\usepackage{amssymb}
\usepackage{amsthm}
\usepackage{amsmath}
\usepackage{multirow}
\usepackage{color}
\usepackage{cite}

\def\beq{\begin{equation}}
\def\eeq{\end{equation}}
\def\bea{\begin{eqnarray}}
\def\eea{\end{eqnarray}}
\newcommand{\beqs}{\begin{subequations}}
\newcommand{\eeqs}{\end{subequations}}

\newcommand{\cref}[1]{Ref.~\cite{#1}}

\newcommand{\bra}[1]{\left<#1\right|}
\newcommand{\ket}[1]{\left|#1\right>}

\newcommand{\hh}{{\ensuremath{I{\kern-2.6pt h}}}}
\newcommand{\bhh}{{\ensuremath{\bar{I{\kern-2.6pt h}}}}}

\def\q2 {q^2}

\def\q2 {q^2}

\def\bra{\langle}
\def\ket{\rangle}

\usepackage[colorlinks=true,allcolors=blue]{hyperref}

\begin{document}

\begin{titlepage}
	
%\vspace*{-15mm}
%\begin{flushright}
%{UT-STPD-21/01}\\
%\end{flushright}
%\vspace*{0.7cm}

\begin{center}
{\Large {\bf Heavier $W$ boson, dark matter and gravitational waves from strings in an SO(10) axion model }
}
\\[12mm]
George Lazarides,$^{1}$~%\footnote{E-mail: \texttt{lazaride@eng.auth.gr}}
Rinku Maji,$^{2}$~
Rishav Roshan,$^{3}$~
Qaisar Shafi$^{4}$~%\footnote{E-mail: \texttt{shafi@bartol.udel.edu}}
\end{center}
\vspace*{0.50cm}
\centerline{$^{1}$ \it
School of Electrical and
Computer Engineering, Faculty of Engineering,
}

\centerline{\it
Aristotle University
of Thessaloniki, Thessaloniki 54124, Greece}
\vspace*{0.2cm}
	\centerline{$^{2}$ \it
		Theoretical Physics Division, Physical Research Laboratory,}
		\centerline{\it  Navarangpura, Ahmedabad 380009, India}
	\vspace*{0.2cm}
	\centerline{$^{3}$ \it
		Department of Physics, Kyungpook National University, Daegu 41566, Korea}
	%\centerline{\it
	%	 Department of Physics, Kyungpook National University, Daegu 41566, Korea}
	\vspace*{0.2cm}
	\centerline{$^{4}$ \it
		Bartol Research Institute, Department of Physics and 
		Astronomy,}
	\centerline{\it
		 University of Delaware, Newark, DE 19716, USA}
	\vspace*{1.20cm}
\begin{abstract}
 Inspired by the recent determination of the $W$-boson mass by the CDF collaboration, we revisit an $SO(10)$ axion model in which a scalar $SU(2)_L$ triplet field with zero hypercharge is known to acquire a non-zero VEV through its mixing with the Standard Model Higgs doublet. The triplet VEV provides a sizable contribution to the $W$ mass, which helps in significantly lowering the $7\sigma$ discrepancy between the Standard Model prediction and the higher CDF value for $m_W$.
We show that the relatively light triplet mass ($\sim (1-50)$ TeV) is compatible with gauge coupling unification and observable proton decay. An unbroken $Z_2$ gauge symmetry, coupled with the presence of two fermionic $10$-plets required to resolve the axion domain wall problem, means that both axions and a stable intermediate mass ($\sim 10^9-10^{10}$ GeV) fermion are plausible dark matter candidates. We also display the gravitational wave spectrum from the intermediate scale topologically stable cosmic strings predicted by the model.

\end{abstract}

%\vspace{1.5cm}
%{\Large\textit{\textbf{\textsl{
%\begin{center}
%The Magnetic Monopole Ninety Years Later
%\end{center}
%}}}}
\end{titlepage}
%%%%%%%%%%%%%%%%%%%%%%%%%%%%%%%%%%%%%%%%%%%%%%%%%%%%%%%
\section{Introduction}
In a recent paper \cite{Lazarides:2021los} largely concerned with the electroweak monopole in Grand Unified Theories (GUTs), it was briefly noted that a specific $SO(10)$ axion model contains a $SU(2)_L$ scalar triplet field with hypercharge $Y=0$ that acquires a non-zero vacuum expectation value (VEV) through its mixing with the Standard Model (SM) Higgs doublet.
It is well-known that this VEV only contributes to the $W$- boson mass, which makes the $SO(10)$ axion model attractive in light of the recent measurement $m_W = 80.4335\pm 0.0094 $ GeV \cite{CDF:2022hxs}. The CDF result is about $7 \sigma$ away from the central value estimated within the SM \cite{Zyla:2020zbs, Lu:2022bgw, deBlas:2022hdk}, and a large number of papers that can explain this deviation has been proposed \cite{Fan:2022dck,Liu:2022jdq,Athron:2022isz,Song:2022xts,Cheung:2022zsb,Endo:2022kiw,Han:2022juu,Ahn:2022xeq,Perez:2022uil,Kawamura:2022uft,Kanemura:2022ahw,Nagao:2022oin,Mondal:2022xdy,Zhang:2022nnh,Carpenter:2022oyg,Popov:2022ldh,Arcadi:2022dmt,Chowdhury:2022moc,Borah:2022obi,Du:2022fqv,Ghorbani:2022vtv,Asadi:2022xiy,Bhaskar:2022vgk,Babu:2022pdn,Ghoshal:2022vzo,Arias-Aragon:2022ats,Sakurai:2022hwh,Gu:2022htv,Batra:2022org,Baek:2022agi,Borah:2022zim,Heeck:2022fvl,Addazi:2022fbj,Cheng:2022aau,Crivellin:2022fdf,Lee:2022gyf,Batra:2022pej,Benbrik:2022dja,Cai:2022cti,Zhou:2022cql,Zhu:2022tpr,Wang:2022dte,Dcruz:2022dao,Li:2022gwc,He:2022zjz,Kim:2022hvh,Evans:2022dgq,Chowdhury:2022dps,Kim:2022zhj,Athron:2022qpo,DiLuzio:2022xns,Heckman:2022the,Chen:2022ocr,Kim:2022xuo,Barman:2022qix}.

In order to realize (better) agreement with the higher value for $m_W$ determined by CDF, the triplet VEV should make a significant contribution to the $W$ mass while maintaining compatibility with the SM $\rho$ parameter. This requires the triplet mass to be of order $10$ TeV or so, and we show how this is achieved in the $SO(10)$ axion model while preserving the unification of the SM gauge couplings. In addition to the axion, the model also contains an intermediate-mass fermion dark matter (DM) candidate whose stability is guaranteed by a discrete $Z_2$ gauge symmetry. This $Z_2$ symmetry is also responsible for the existence of topologically stable cosmic strings.

The plan of the paper is as follows. In Section~\ref{sec:model} we summarize the salient features of the model including the symmetry breaking pattern and the realization of higher $m_W$ compared to the SM. Section~\ref{sec:unific} deals with gauge coupling unification and implications for proton decay. Section~\ref{sec:DM} discusses the DM candidates consisting of axions and intermediate-mass neutral fermions. The gravitational wave (GW) spectrum from the intermediate scale cosmic strings is discussed in Section~\ref{sec:GWs}, and we conclude with a summary in Section~\ref{sec:summary}.
%%%%%%%%%%%%%%%%%%%%%%%%%%
\section{The Model}\label{sec:model}
In this section, we briefly outline the salient features of the $SO(10)$ axion model and refer the reader to
Refs.~\cite{Holman:1982tb, Lazarides:2020frf} for additional details. To start with, we first describe the particle content of the setup and then present all the relevant interactions of these particles. We denote the fermion multiplets present in the model as 
\begin{align}
\psi^{(i)}_{16}(1) \ \ (i = 1,2,3)  , \ \  \psi^{(\alpha)}_{10}(-2) \quad (\alpha = 1,2) ,
\end{align}
and the scalar multiplets as
%and the scalar sector consists of the following multiplets
\begin{align}
\phi_{10}(-2), \ \ \phi_{45}(4), \ \ \phi_{126}(2), \ \ \phi_{210}(0) .
\end{align}
Here, the subscripts refer to the dimension of the representations under $SO(10)$, the Peccei-Quinn (PQ) charges ($Q_{\rm PQ}$) are quoted within parentheses, and $i$ and $\alpha$ are the generation indices for the fermions. With the knowledge of the particle spectrum and symmetries of the model, next, we present all the relevant interactions involving these fields. The Yukawa couplings are
\begin{align}
\psi^{(i)}_{16}\psi^{(j)}_{16}\phi_{10}, \ \ \psi^{(i)}_{16}\psi^{(j)}_{16}\phi_{126}^\dagger, \ \ \psi^{(1)}_{10}\psi^{(2)}_{10}\phi_{45} ,
\end{align} 
and the scalar couplings include 
\begin{align}\label{eq:scalar_coupling}
 \phi_{210}\phi_{126}^\dagger\phi_{126}^\dagger\phi_{45}, \ \ \phi_{210}\phi_{126}^\dagger\phi_{10}\phi_{45}, \ \ \phi_{210}\phi_{126}\phi_{10} .
\end{align}
The SM fermions of each family, accompanied by a SM singlet right-handed neutrino, reside in the $16$-dimensional representation of $SO(10)$. In addition, two generations of fermionic 10-plets are included to overcome the axion domain wall problem \cite{Holman:1982tb, Lazarides:2020frf}. The electroweak sector of the model contains the SM Higgs doublet, which is a linear combination of the two $SU(2)_L$ doublets from $\phi_{10}$ and two doublets from $\phi_{126}$. The three remaining scalar doublets obtain masses of order $M_{II}$ \cite{Babu:2015bna}.

 At this stage some remarks about the so-called quality problem in axion models are in order. If one makes the rather arbitrary assumption that Planck scale suppressed operators are present in axion models, it then follows that they must not be permitted to spoil the axion resolution of the strong CP problem. We are therefore lead to the conclusion that the coefficients accompanying the potentially dangerous operators, dimension five (e.g., $\phi_{45}^4\phi_{210}$, $\phi_{45}^\dagger\phi_{45}^3\phi_{210}$) and some higher ones in our case, must be adequately suppressed. Clearly, such operators do not arise in the renormalizable $SO(10)$ framework, and their occurrence in the presence of gravity has not been convincingly demonstrated. Indeed, it has been suggested that wormhole tunneling may give rise to $U(1)_{\rm PQ}$ symmetry violating effects with exponentially suppressed coefficients, and they only become important for $f_a \gtrsim 10^{17}$ GeV \cite{Giddings:1987cg,Lee:1988ge,Alvey:2020nyh}. In our model the axion decay constant $f_a$ is, of course, orders of magnitude smaller than $10^{17}$ GeV and the problem is therefore avoided.

 Be that as it may, perhaps a more elegant approach for resolving the axion quality problem is to assume that a suitable discrete gauge symmetry effectively behaves as $U(1)_{\rm PQ}$. Discrete gauge symmetries routinely arise from the four dimensional compactification of higher dimensional superstring theories, and the first examples based on this idea have been discussed in Ref.~\cite{Lazarides:1985bj}.

 Finally, as shown in Ref.~\cite{DiLuzio:2020qio}, it is possible that $U(1)_{\rm PQ}$ may appear as an accidental symmetry in $SO(10)$ models supplemented by a suitable continuous gauge symmetry. In this case too the axion quality problem is suitably ameliorated.

 For definiteness, we employ a specific symmetry breaking pattern of $SO(10)$ shown in Eq.~(\ref{eq:breaking-chain}) which, among other things, also allows a light $SU(2)_L$ scalar triplet from $\phi_{45}$ that remains compatible with the unification of the SM gauge couplings. Note that the induced VEV of the scalar triplet arises from the coupling $\phi_{10}\phi_{10}\phi_{45}$.
%%%%%%%%%%%%%%%%%%%%%%%%%%%%%%%%%%%%%%%%%
\begin{align}\label{eq:breaking-chain}
& SO(10)\times U(1)_{\rm PQ}\xrightarrow[M_U]{\left<210(0)\right>} \nonumber \\
& SU(2)_L\otimes SU(2)_R\otimes SU(4)_C\times U(1)_{\rm PQ} \xrightarrow[M_I]{\left<(1,1,15)\in 210(0)\right>} \nonumber \\
& SU(2)_L\otimes SU(2)_R\otimes SU(3)_C \otimes U(1)_{B-L}\times U(1)_{\rm PQ}\xrightarrow[M_{II}]{\left< (1,3,1,-2)\in (1,3,10)\in \overline{126}(-2)\right>} \nonumber \\
& SU(3)_C \otimes SU(2)_L\otimes U(1)_Y\otimes\mathbb{Z}_2 \times U(1)'_{\rm PQ} \xrightarrow[f_a]{\left<\lbrace (1,1,0)\in(1,3,1)+(1,1,0)\in(1,1,15)\rbrace\in 45(4)\right>}  \nonumber \\
& SU(3)_C \otimes SU(2)_L\otimes U(1)_Y\otimes\mathbb{Z}_2 \xrightarrow[m_W]{\left< (1,2,\pm \frac{1}{2})\in 10(-2)\right>} SU(3)_C \otimes U(1)_Q\otimes\mathbb{Z}_2 .
\end{align}
%%%%%%%%%%%%%%%%%%%%%%%%%%%%%%%%%%%%%%%%%%%%%

We employ two-loop renormalization group equations (RGEs) to estimate the GUT scale ($M_U$) and the two gauge symmetry breaking intermediate-scales $M_I$ and $M_{II}$. We find it instructive and useful to distinguish the two latter scales from the axion symmetry breaking scale $f_a$ ($\leq M_{II}$). The remnant anomalous global symmetry after $M_{II}$ is $U(1)_{\rm PQ}'$, which is generated by $Q'_{\rm PQ} = 5Q_{\rm PQ} - 3(B-L) + 4 T_R^3$, where $T_R^3$ is the diagonal generator of $SU(2)_R$. The $U(1)_{\rm PQ}'$ symmetry is broken by the VEV of $(1,1,15)$ and $(1,3,1)$ in $45(4)$ at the scale $f_a$. The fermions from $\psi_{10}$ acquire masses during this symmetry breaking which we assume are all of the same order of magnitude, $m_{\rm DM} = y_{45}\left<\phi_{45}\right>$. The lightest neutral fermion from the 10-plets along with the axion can account for the observed dark matter relic density of the universe \cite{Planck:2018vyg} (see Ref.~\cite{Lazarides:2020frf} for details). 

At this stage, it is important to point out that the above breaking chain allows a light $SU(2)_L$ scalar triplet from $\phi_{45}$ that is compatible with the unification of the gauge couplings. We now shed light on this scalar triplet and describe its role in raising the $W$-boson mass above the SM prediction as suggested by the CDF result. We can write the scalar triplet interaction that arises from the term $\phi_{10}\phi_{10}\phi_{45} + {\rm h.c.}$  of Eq.~(\ref{eq:scalar_coupling}) as
\begin{align}\label{eq:hhT}
-\lambda m_T {H_{10}^u}^{\rm T} i\sigma_2 T^i \sigma_i H_{10}^d + {\rm h.c.}
\end{align}
Here $T^i\equiv (1,3,0)$ is the complex triplet scalar from $\phi_{45}$, and $H_{10}^u(\equiv (1,2,\frac{1}{2}))\oplus H_{10}^d(\equiv (1,2,-\frac{1}{2}))$ arise from the bi-doublet $(2,2,1)\in \phi_{10}$.  As a result of electroweak breaking and the presence of this term, a non-zero VEV is induced for the scalar triplet 
\begin{align}
v_T=\sqrt{2} \lambda v_{10}^u v_{10}^d/m_T ,
\end{align}
where $\left<T^3\right>=v_T/\sqrt{2}$ and $\left<H_{10}^d\right>=v_{10}^d/\sqrt{2}$,
$\left<H_{10}^u\right>=v_{10}^u/\sqrt{2}$. The induced triplet VEV modifies the $W$-boson mass such that
\begin{align}\label{eq:w-mass}
m_W^2 = \frac{g^2}{4}\left(v_{\rm SM}^2+4v_T^2\right), \quad {\rm with} \; g_{2L}(m_Z)\equiv g,
\end{align}
 and $v_{\rm SM} =246$ GeV is the SM VEV. This means (as also shown in Eq.~(11) of Ref.~\cite{Popov:2022ldh}) that the electroweak mixing angle and $Z$-boson mass remain unaltered, and the change in the $\rho$ parameter is solely due to the $W$-mass anomaly. Following  Eq.~(\ref{eq:w-mass}), the $\rho$-parameter can be expressed as
\begin{align}\label{eq:rho}
 \rho = 1 + 4(v_T/v_{\rm SM})^2.
 \end{align}
The experimental value of $\rho$ in this case is $1.00219 \pm 0.00044$ (see
Ref.~\cite{Popov:2022ldh}) which, in turn, implies that the central value of the triplet VEV is $v_T=5.7561$ GeV.
%%%%%%%%%%%%%%%%%%%%%%%%%%%%%%%%%%%%%%%%%%%%%%%%%%%%

Before closing this section, we would also like to make a few remarks about the topological defects in this model. The $SO(10)$ breaking at $M_U$ to $SU(2)_L\otimes SU(2)_R\otimes SU(4)_C$ yields a topologically stable monopole that subsequently turns into a superheavy monopole carrying a single unit of Dirac magnetic charge as well as some color magnetic charge. This monopole is inflated away within a suitable inflationary setting as shown, for instance, in Refs.~\cite{Senoguz:2015lba,Chakrabortty:2020otp}. The second breaking yields a stable monopole significantly lighter than $M_U$ that carries two quanta of Dirac charge as well as color charge. Depending on the magnitude of the symmetry breaking scale $M_I$ versus $H_{\rm inf}$, the Hubble parameter during inflation, this monopole with mass $\sim 10 M_I$ may be present in our galaxy at an observable level.

As previously mentioned the unbroken $Z_2$ gauge symmetry implies the presence of topologically stable cosmic strings whose mass scale is determined by the second intermediate scale $M_{II}$. We will discuss these strings and their gravitational wave emission in Section~\ref{sec:GWs}. Finally, for completeness let us note that the axion strings in this model appear after inflation and form a string-wall system at the QCD phase transition. The strings are superconducting and the loops emit axions and perhaps even the intermediate scale fermion dark matter.

%%%%%%%%%%%%%%%%%%%%%%%%%%%%%%%%%%%%%%%%%%%
\section{Unification Solutions}\label{sec:unific}
We aim to obtain unification solutions compatible with the electroweak observables (see Table~\ref{tab:ewpo}) in terms of the unified gauge coupling ($g_U$), intermediate scales ($M_I$ and $M_{II}$), and unification scale ($M_U$) for different choices of $m_{\rm DM}$ and triplet scalar mass ($m_T$). 
%%%%%%%%%%%%%%%%%%%%%%%%%%%%%%%%%%%%%%%%
\begin{table}[htbp]
 	\begin{center}
 		\begin{tabular}{|c|c|}
 			\hline
 			 $Z$-boson mass, $m_Z$ & $91.1876(21)$ GeV \\
 			\hline
 			Strong fine structure constant, $\alpha_{3C}$ & $0.1179(10)$ \\
 			\hline
 			Fermi coupling constant, $G_F$ & $1.1663787(6)\times 10^{-5} \ \rm{GeV}^{-2}$ \\
 			\hline
 			Electroweak mixing angle, $\sin^2{\theta_W}$ & $0.23121(4)$ \\
 			\hline
 		\end{tabular}
 		\caption{Electroweak observables at $m_Z$ \cite{Zyla:2020zbs}.}\label{tab:ewpo}
 	\end{center}
 \end{table}
%%%%%%%%%%%%%%%%%%%%%%%%%%%%%%%%%%%%%%%%%%%%%%%%
We minimize the $\chi^2$ defined at $m_Z$ and given by
\begin{equation}
\chi^2 = {\sum_{i=1}^3 \frac{\left(g_i^2 - g_{i,\mathrm{exp}}^2\right)^2}{\sigma^2_{g^2_{i,\mathrm{exp}}}}},
\end{equation}
where $g_i$ ($i=Y, 2L, 3C$) are the SM gauge couplings at $m_Z$ obtained through the RGEs starting from the unified gauge coupling at the unification scale and $g_{i,\mathrm{exp}}$ are their experimental values.
We compute the $\beta$-coefficients as outlined in Refs.~\cite{Jones:1981we,Chakrabortty:2017mgi,Chakrabortty:2019fov}. The one- and two-loop $\beta$-coefficients governing the renormalization group evolution of the gauge couplings at different stages are given in Table~\ref{tab:RGE_beta_coef}. 
%%%%%%%%%%%%%%%%%%%%%%%%%%%%%%%%%%%%%%%%
\begin{table}[htbp]
\begin{center}
\begin{tabular}{| c || c |}
\hline
 $\mathcal{G}_{2_L2_R4_C}\times U(1)_{PQ}$ & $\mathcal{G}_{2_L2_R3_C1_{B-L}}\times U(1)_{PQ}$  \\
 \hline
$\begin{pmatrix}
4 \\ 
\frac{32}{3} \\ 
\frac{5}{3}
\end{pmatrix}$ , 
$\begin{pmatrix}
108 & 51 & \frac{525}{2} \\ 
51 & \frac{884}{3} & \frac{1245}{2} \\ 
\frac{105}{2} & \frac{249}{2} & \frac{3551}{6}
\end{pmatrix}$ & $\begin{pmatrix}
-\frac{2}{3} \\ 
0 \\ 
-\frac{17}{3} \\ 
\frac{41}{6}
\end{pmatrix}$ , 
$\begin{pmatrix}
\frac{142}{3} & 9 & 12 & \frac{3}{2} \\ 
9 & 66 & 12 & \frac{27}{2} \\ 
\frac{9}{2} & \frac{9}{2} & -\frac{2}{3} & \frac{7}{6} \\ 
\frac{9}{2} & \frac{81}{2} & \frac{28}{3} & \frac{187}{6}
\end{pmatrix}$ \\
\hline\hline
$\mathcal{G}_{3_C2_L1_Y\mathbb{Z}_2} \times U(1)'_{PQ}$ & $\mathcal{G}_{3_C2_L1_Y\mathbb{Z}_2}$ (Triplet) \\
\hline
 $\begin{pmatrix}
-\frac{17}{3} \\ 
-\frac{7}{6} \\ 
\frac{163}{30}
\end{pmatrix}$ , 
$\begin{pmatrix}
-\frac{2}{3} & \frac{9}{2} & \frac{41}{30} \\ 
12 & \frac{245}{6} & \frac{3}{2} \\ 
\frac{164}{15} & \frac{9}{2} & \frac{667}{150}
\end{pmatrix}$ & $\begin{pmatrix}
-7 \\ 
-\frac{5}{2} \\ 
\frac{41}{10}
\end{pmatrix}$ , 
$\begin{pmatrix}
-26 & \frac{9}{2} & \frac{11}{10} \\ 
12 & \frac{49}{2} & \frac{9}{10} \\ 
\frac{44}{5} & \frac{27}{10} & \frac{199}{50}
\end{pmatrix}$\\
\hline
\end{tabular}
\caption{One- and two-loop beta coefficients for the renormalization group evolution of the gauge couplings at different stages of gauge symmetry.}\label{tab:RGE_beta_coef}
\end{center}
\end{table}
%%%%%%%%%%%%%%%%%%%%%%%%%%%%%%%%%%%%%%%%%

\begin{figure}[htbp]
\centering
\includegraphics[width=0.7\linewidth]{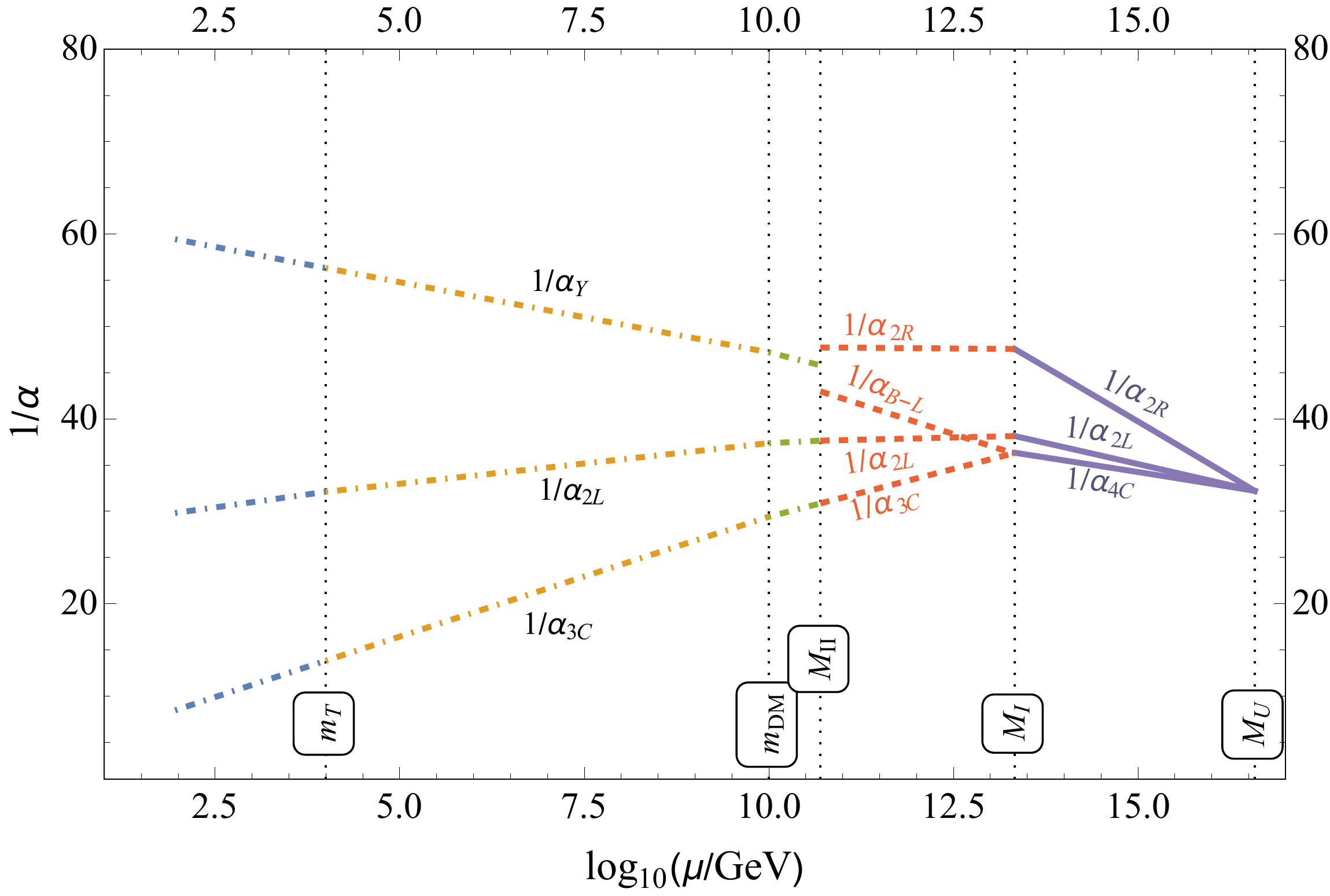}
\caption{Renormalization group evolution of the gauge couplings for a unification solution with $m_T=10$ TeV, $m_{\rm DM}=10^{10}$ GeV, $M_{II}=5.0\times 10^{10}$ GeV, $M_I=2.15\times 10^{13}$ GeV, $M_U=3.8\times 10^{16}$ GeV, and $g_U=0.624$.}\label{fig:rge-run}
\end{figure}
We show the RGE running of the gauge couplings %at different stages of symmetry 
in Fig.~\ref{fig:rge-run} for a unification solution with $m_T=10$ TeV, $m_{\rm DM}=10^{10}$ GeV, $M_{II}=5.0\times 10^{10}$ GeV, $M_I=2.15\times 10^{13}$ GeV, $M_U=3.8\times 10^{16}$ GeV, and $g_U=0.624$.
%%%%%%%%%%%%%%%%%%%%%%%%%%%%%%%%%%%%%%%%%%%%%%%%%%%%%%%%
\begin{table}
\centering
\begin{tabular}{|c|c|c|c|c|c|}
\hline
$m_{\rm DM}$ & $m_T$(TeV) & $\log_{10}\left(\frac{M_U}{\rm GeV}\right)$ & $\log_{10}\left(\frac{M_I}{\rm GeV}\right)$ & $\log_{10}\left(\frac{M_{II}}{\rm GeV}\right)$ & $g_U$ \\
\hline
\multirow{4}{*}{\rotatebox[origin=c]{90}{$10^9$ GeV}} & 1 & \{17.75,15.65\} & \{12.57,13.75\} & \{9.0,12.3\} & \{0.679,0.605\} \\
& 5 & \{17.68,15.63\} & \{12.71,13.88\} & \{9.0,12.2\} & \{0.671,0.600\} \\
& 10 & \{17.65,15.66\} & \{12.77,13.91\} & \{9.0,12.1\} & \{0.668,0.600\} \\
& 50 & \{17.58,15.63\} & \{12.91,14.04\} & \{9.0,12.0\} & \{0.661,0.596\} \\
\hline
\multirow{4}{*}{\rotatebox[origin=c]{90}{$10^{10}$ GeV}} & 1 & \{17.14,15.65\} & \{12.86,13.74\} & \{10.0,12.3\} & \{0.649,0.600\} \\
& 5 & \{17.07,15.62\} & \{13.01,13.87\} & \{10.0,12.2\} & \{0.642,0.596\} \\
& 10 & \{17.03,15.65\} & \{13.07,13.90\} & \{10.0,12.1\} & \{0.639,0.596\} \\
& 50 & \{16.96,15.63\} & \{13.23,14.03\} & \{10.0,12.0\} & \{0.633,0.592\} \\
\hline
\end{tabular}
\caption{Unification solutions for the unification scale $M_U$, intermediate scales $M_I$ and $M_{II}$, and unified coupling $g_U$ for different choices of $m_{\rm DM}$ and $m_T$.}\label{tab:unific-soln}
\end{table}

%%%%%%%%%%%%%%%%%%%%%%%%%%%%%%%%%%%%%%%%%%%
\begin{figure}[htbp!]
\centering
\subfloat[$m_{\mathrm{DM}}=10^9$ GeV.]{\includegraphics[width=0.7\linewidth]{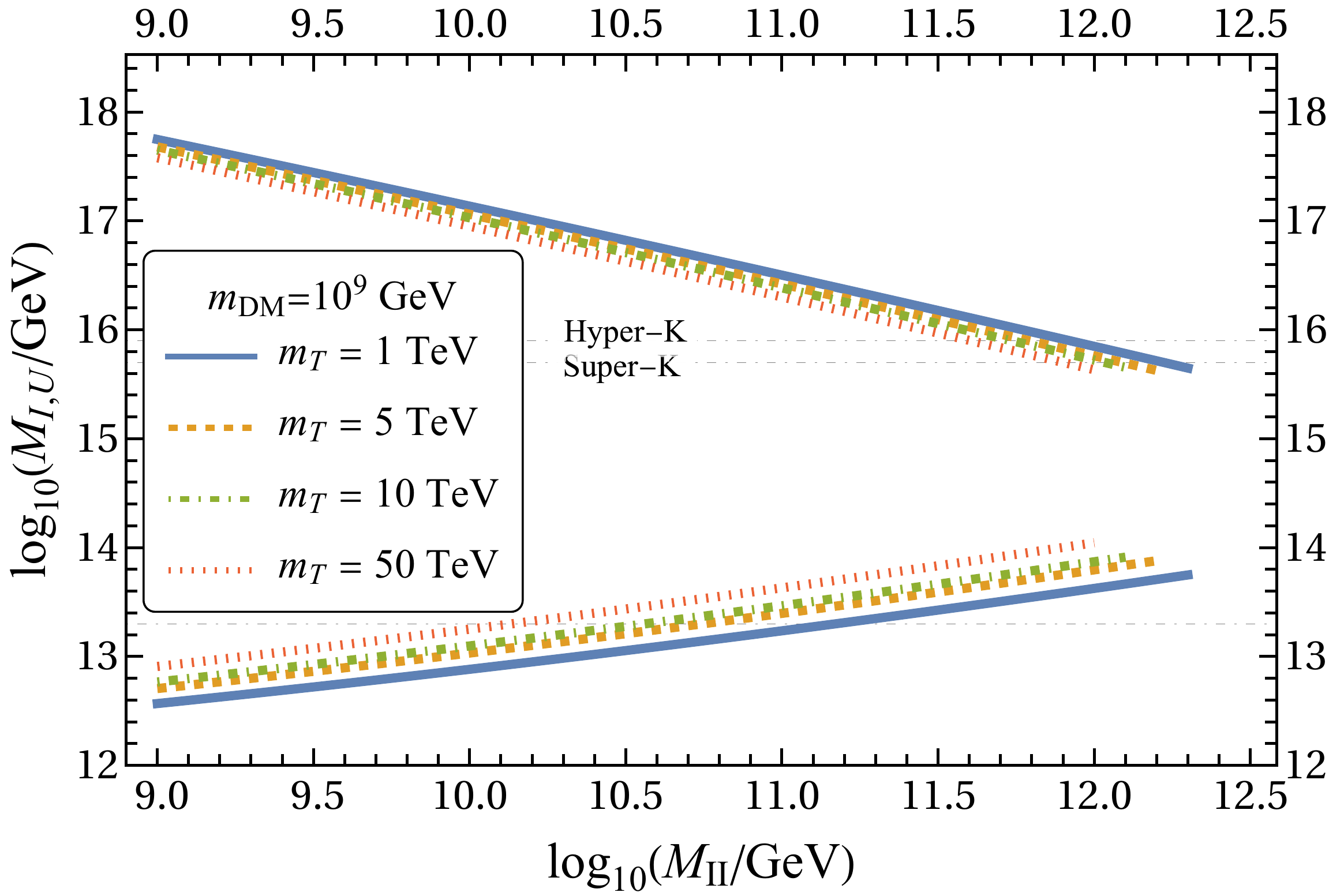}} \\
\subfloat[ $m_{\mathrm{DM}}=10^{10}$ GeV.]{\includegraphics[width=0.7\linewidth]{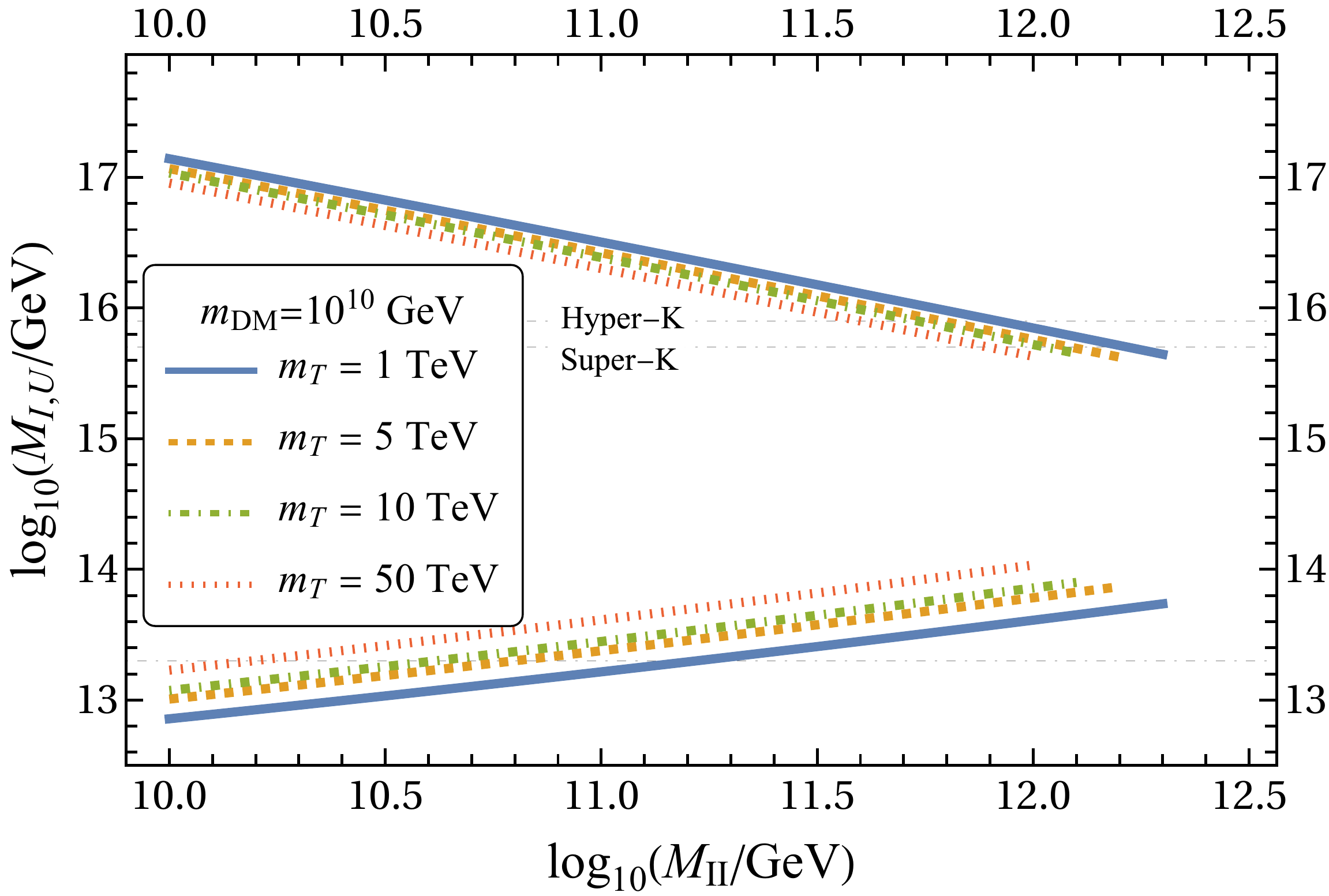}}
\caption{Variation of unification scale ($M_U$) and first intermediate breaking scale ($M_I$) with the second intermediate breaking scale ($M_{II}$) for different choices of the triplet scalar mass ($m_T$) and dark matter mass ($m_{\rm DM}$). The horizontal dot-dashed lines at $\log_{10} \left( M_U/\mathrm{GeV}\right)$ equal to $15.7$ and $15.9$ are the lower bound on $M_U$ from the Super-Kamiokande experiment and the sensitivity of the proposed Hyper-Kamiokande experiment respectively. The horizontal dot-dashed line at $\log_{10} \left( M_I/\mathrm{GeV}\right) = 13.3$ is the lower bound from the MACRO experiment within the inflationary scenario driven by the Coleman-Weinberg potential of a real GUT singlet \cite{Chakrabortty:2020otp}.}\label{fig:plt_CDF}
\end{figure}
%%%%%%%%%%%%%%%%%%%%%%%%%%%%%%%%%%%%%%%%%%%%%

Next, in Table~\ref{tab:unific-soln} we show the unification solutions for two typical values $10^9$ GeV and $10^{10}$ GeV of $m_{\rm DM}$ with $m_T = \lbrace 1, 5, 10, 50 \rbrace$ TeV for each value of $m_{\rm DM}$. In Fig.~\ref{fig:plt_CDF}, we have plotted the unification scale ($M_U$) and first intermediate scale ($M_I$) as functions of the second intermediate scale ($M_{II}$) for different choices of $m_{\rm DM}$ and $m_T$. 

The non-observation of proton decay in the Super-Kamiokande (Super-K) experiment has pushed the partial lifetime bound for the decay channel $p\to e^+\pi^0$ to be above $2.4\times 10^{34}$ yrs \cite{Super-Kamiokande:2020wjk}, which constrains the unification scale $M_U \gtrsim 5.3\times 10^{15}$ GeV. On the other hand, the Hyper-Kamiokade (Hyper-K) experiment has $3\sigma$ discovery potential to probe the channel $p\to e^+\pi^0$ with partial lifetime  $10^{35}$ yrs \cite{Dealtry:2019ldr} which corresponds to $M_U \simeq 7.5\times 10^{15}$ GeV. We have indicated the Super-K lower limit and the Hyper-K sensitivity in Fig.~\ref{fig:plt_CDF}. There are unification solutions that are compatible with the Super-K bound and a part of them will be probed by the Hyper-K experiment as can be seen in Fig.~\ref{fig:plt_CDF}. The monopoles produced during the symmetry breaking at the scale $M_I$ should be partially inflated to satisfy the lower bound on the monopole flux $2.8\times 10^{-16} \ \mathrm{cm}^{-2} \mathrm{s}^{-1} \mathrm{sr}^{-1} $ \cite{Ambrosio:2002qq}. In the inflationary scenario driven by the Coleman-Weinberg potential of a real GUT singlet, the monopoles undergo a sufficient number of $e$-foldings to comply with the MACRO bound for $M_I \gtrsim 2\times 10^{13}$ GeV \cite{Chakrabortty:2020otp} which is compatible with a good part of the unification solutions as shown in Fig.~\ref{fig:plt_CDF}. 

%%%%%%%%%%%%%%%%%%%%%%%%%%%%%%

%%%%%%%%%%%%%%%%%%%%%%%%%%%%%%%%%%%%%%%%%%%%%%%%%%%%%%%%%%%%%%%%%%
\section{Axion and Intermediate Scale Fermion Dark Matter}
\label{sec:DM}

\begin{figure}[htb!]
	\centering
	\includegraphics[width=0.7\linewidth]{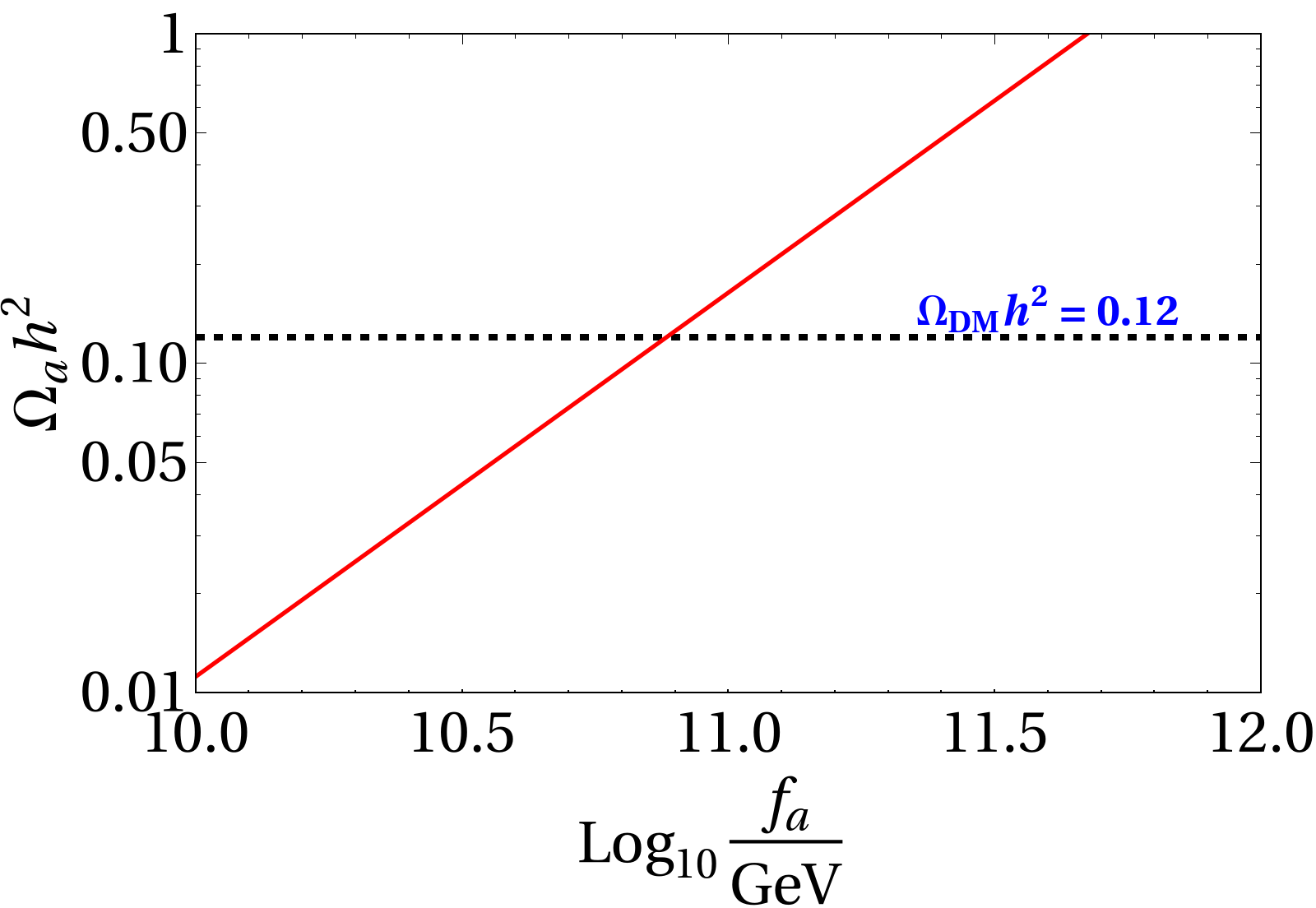}
	\caption{Variation of the axion relic density with the axion decay constant $f_a$. The black dashed line corresponds to $\Omega_{\text{DM}}h^2=0.12$.}
	\label{relic_axion}
\end{figure}

In this section we investigate the scenario of axion and the lightest neutral component of the 10-plet as DM candidates in the model such that
\bea
\Omega_\text{Total}h^2=\Omega_ah^2 + \Omega_{10}h^2,
\label{total_relic}
\eea
where $\Omega_ah^2$ and $\Omega_{10}h^2$ denote the axion and fermion relic densities respectively. It is interesting to point out that in this model axions can be produced by two different mechanisms, namely (a) the misalignment mechanism~\cite{Preskill:1982cy,Abbott:1982af,Stecker:1982ws,Visinelli:2009zm} and (b) the decay of axionic strings~\cite{Hagmann:2000ja,Visinelli:2009zm}.
The relic axion abundance produced by the misalignment mechanism is expressed as~\cite{Visinelli:2009zm}
\bea
\Omega_a^{mis}h^2\simeq 0.236 \bigg(\frac{f_a}{10^{12}~\text{GeV}}\bigg)^{7/6}\bra\theta^2f(\theta)\ket,
\label{relic_axion_exp_mis}
\eea
where $\theta$ denotes the misalignment angle that lies in the interval $[-\pi,\pi]$ \cite{Dimopoulos:2003ii}. The function $f(\theta)$ contains the anharmonicity of the axion potential, and $\bra\theta^2f(\theta)\ket$ evaluated in the interval $[-\pi,\pi]$ turns out to be around 8.77~\cite{Visinelli:2009zm}. As previously discussed in Section~\ref{sec:model}, the decay of $U(1)_\text{PQ}$ strings also contributes significantly in the production of axions and hence cannot be ignored. This contribution to the relic density can be expressed as~\cite{Visinelli:2009zm}
\bea
\Omega_a^{str}h^2\simeq 0.34 \bigg(\frac{f_a}{10^{12}~\text{GeV}}\bigg)^{7/6}.
\label{relic_axion_exp_str}
\eea
The total axion relic density is thus given by
\bea
\Omega_ah^2=\Omega_a^{mis}h^2+\Omega_a^{str}h^2\simeq 2.41 \bigg(\frac{f_a}{10^{12}~\text{GeV}}\bigg)^{7/6}.
\label{relic_axion_total}
\eea
In Fig.~\ref{relic_axion}, we show how $\Omega_ah^2$ varies with $f_a$, with the black dashed line denoting the Planck limit~\cite{Planck:2018vyg} on the relic DM abundance. With $f_a\simeq8\times10^{10}$ GeV, the axion saturates the observed DM relic density. 

For $f_a$ smaller than $8\times10^{10}$ GeV, the contribution from the fermionic DM component should be taken into account. We do not aim to discuss the production mechanism of the fermion DM but provide, instead, a rough analytical estimate of its abundance ($Y_\text{DM}=n_\text{DM}/s$). The relic density of the fermion DM can be expressed as
\bea
\Omega_\text{Total}h^2-\Omega_ah^2 =\frac{m_\text{DM}Y_\text{DM}s_0}{\rho_c} ,
\label{relic_fermion}
\eea
where $s_0\simeq 2890 ~\text{cm}^{-3}$ is the present entropy density and $\rho_c\simeq1.05\times10^{-5}~\text{GeV cm}^{-3}$ is the present day critical density. Using Eq.~(\ref{relic_fermion}),
we find that
\bea
Y_\text{DM}\simeq4.36\times10^{-10}(\Omega_\text{Total}h^2-\Omega_ah^2)\bigg(\frac{\text{GeV}}{m_\text{DM}}\bigg).
\label{yield}
\eea
%%%%%%%%%%%%%%%%%%%%%%%%%%%%%%%%%%%%%%%%%%%%
\begin{figure}[htb!]
	\centering
	\includegraphics[width=0.7\linewidth]{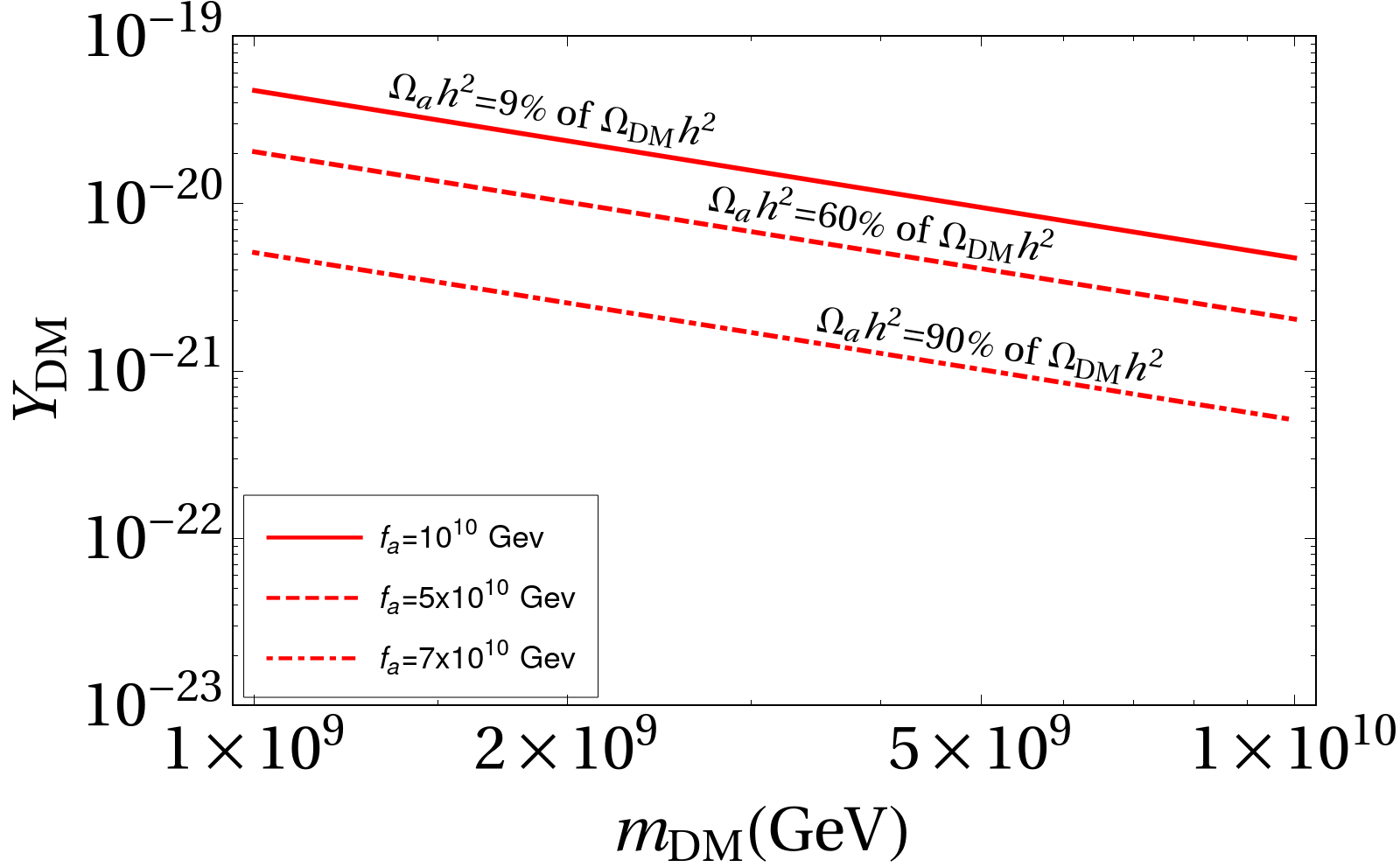}
	\caption{Variation of the fermion DM yield versus its mass for three different values of $f_a$: $10^{10}$ GeV (red solid), $5\times10^{10}$ GeV (red dashed) and $7\times10^{10}$ GeV (red dot-dashed).}
	\label{yield_DM}
\end{figure} 
%%%%%%%%%%%%%%%%%%%%%%%%%%%%%%%%%%%%%%%%%%%
In Fig.~\ref{yield_DM}, we show the variation of the asymptotic yield of the fermion DM %(that contributes significantly to the total DM relic density) 
with its mass for three different values of the axion decay constant $f_a$. The solid red line that corresponding to $f_{a}=10^{10}$ GeV suggests that around $91\%$ of the total relic density of the DM is composed of intermediate mass scale fermions, with the remaining $9\%$ coming from axions. As expected, making $f_a$ larger increases the axion contribution to the total DM relic density (see Fig.~\ref{relic_axion}), and the corresponding contribution from the fermion DM has to be reduced. This can be seen from the red dashed ($f_{a}=5\times10^{10}$ GeV) and red dot-dashed ($f_{a}=7\times10^{10}$ GeV) lines in Fig.~\ref{yield_DM}. 

%%%%%%%%%%%%%%%%%%%%%%%%%%%%%%%%%%%%%%%%%%%%%%%%%%%%%%%%%%%%%%%%%%
\section{Gravitational Waves from Cosmic String Loops}\label{sec:GWs}
The spontaneous symmetry breaking at $M_{II}$ generates local cosmic strings which are topologically stable. The dimensionless tension of the strings is given by
\begin{align}
G\mu = \frac{1}{8}\left(\frac{M_{II}}{m_{\rm Pl}}\right)^2 ,
\end{align}
where $G$ is Newton's gravitational constant and $m_{\rm Pl}$ is the reduced Planck mass.
The strings intercommute and form loops that decay by emitting GWs. We estimate the gravitational wave spectra following the burst method described in Refs.~\cite{Olmez:2010bi,Cui:2019kkd,Auclair:2019wcv}. To this end, we need the loop distribution function $n(l,t)$ (the number density of loops per unit loop length $l$) at the time of GW emission. This is given in the different cosmic epochs in Refs.~\cite{Blanco-Pillado:2013qja,Blanco-Pillado:2017oxo} (also see the supplemental material of Ref.~\cite{LIGOScientific:2021nrg}).

In the radiation dominated universe, we have
\begin{align}\label{eq:n-loop-rad}
n_r(l,t) = \frac{0.18}{t^{3/2}(l+\Gamma G\mu t)^{5/2}}\Theta(0.1t-l)  ,
\end{align}
where $\Gamma \simeq 50$.
In the matter dominated universe there are two contribution. For the loops that are remnants from the radiation era
\begin{align}\label{eq:n-loop-radmat}
n_{rm}(l,t) = \frac{0.18t_{eq}^{1/2}}{t^2(l+\Gamma G\mu t)^2}\Theta(0.18t_{eq}-l-\Gamma G\mu(t-t_{eq})) \ ,
\end{align}
where $t_{eq}$ is the equidensity time,
and for the loops that are produced during the matter dominated era
\begin{align}\label{eq:n-loop-mat}
n_{m}(l,t) = \frac{0.27-0.45(l/t)^{0.31}}{t^2(l+\Gamma G\mu t)^2}\Theta(0.18t-l)\Theta(l+\Gamma G\mu(t-t_{eq})-0.18t_{eq}) \ .
\end{align}
Assuming cusp domination, the waveform at frequency $f$ and redshift $z$ is given by
\begin{align}
h(f,l,z) = g_{1c}\frac{G\mu \, l^{2/3}}{(1+z)^{1/3}r(z)}f^{-4/3} ,
\end{align}
where $g_{1c}\simeq 0.85$ \cite{LIGOScientific:2021nrg} and $r$ is the proper distance
\begin{align}
r(z)=\int_0^z\frac{dz'}{H(z')} ,
\end{align}
with $H$ being the Hubble parameter.
For the burst rate per unit space-time volume we have
\begin{align}
\frac{d^2R}{dz \, dl} = N_c H_0^{-3}\phi_V(z) \frac{2n(l,t(z))}{l(1+z)}\left( \frac{\theta_m(f,l,z)}{2}\right)^2\Theta(1-\theta_m) ,
\end{align}
where $H_0$ is the present value of the Hubble parameter,
\begin{align}
\theta_m(f,l,z) = \left[\frac{\sqrt{3}}{4}(1+z)fl\right]^{-1/3}
\end{align}
is the beam opening angle, and
\begin{align}
\phi_V(z) = \frac{4\pi H_0^3r^2}{(1+z)^3H(z)} \ .
\end{align}
 We have taken $N_c=2.13$ as in Ref.~\cite{Cui:2019kkd}.
 
The GW background is given by
\begin{align}\label{eq:GWs-Omega-cusps}
\Omega_{GW}(f) = \frac{4\pi^2}{3H_0^2}f^3\int_{z_*}^{z(t_F)}dz \int dl \, h^2(f,l,z)\frac{d^2R}{dz \, dl} \ ,
\end{align}
where $t_F$ is the time when loop formation starts and the lower limit $z_*$ in the integral in Eq.~(\ref{eq:GWs-Omega-cusps}) leaves out the infrequent bursts from the stochastic background \cite{Cui:2019kkd} so that
\begin{align}
\int_0^{z_*} dz  \int dl \,\frac{d^2R}{dz dl} = f .
\end{align}
We have taken the integration limit on $l$ to be from $0$ to $2t$ ($3t$) for the radiation (matter) domination. The various Heaviside $\Theta$ functions will anyway control the upper and lower integration limits during numerical evaluations. 
%%%%%%%%%%%%%%%%%%%%%%%%%%%%%%%%%%%%%% 
\begin{figure}[htbp]
\begin{center}
\includegraphics[width=0.7\linewidth]{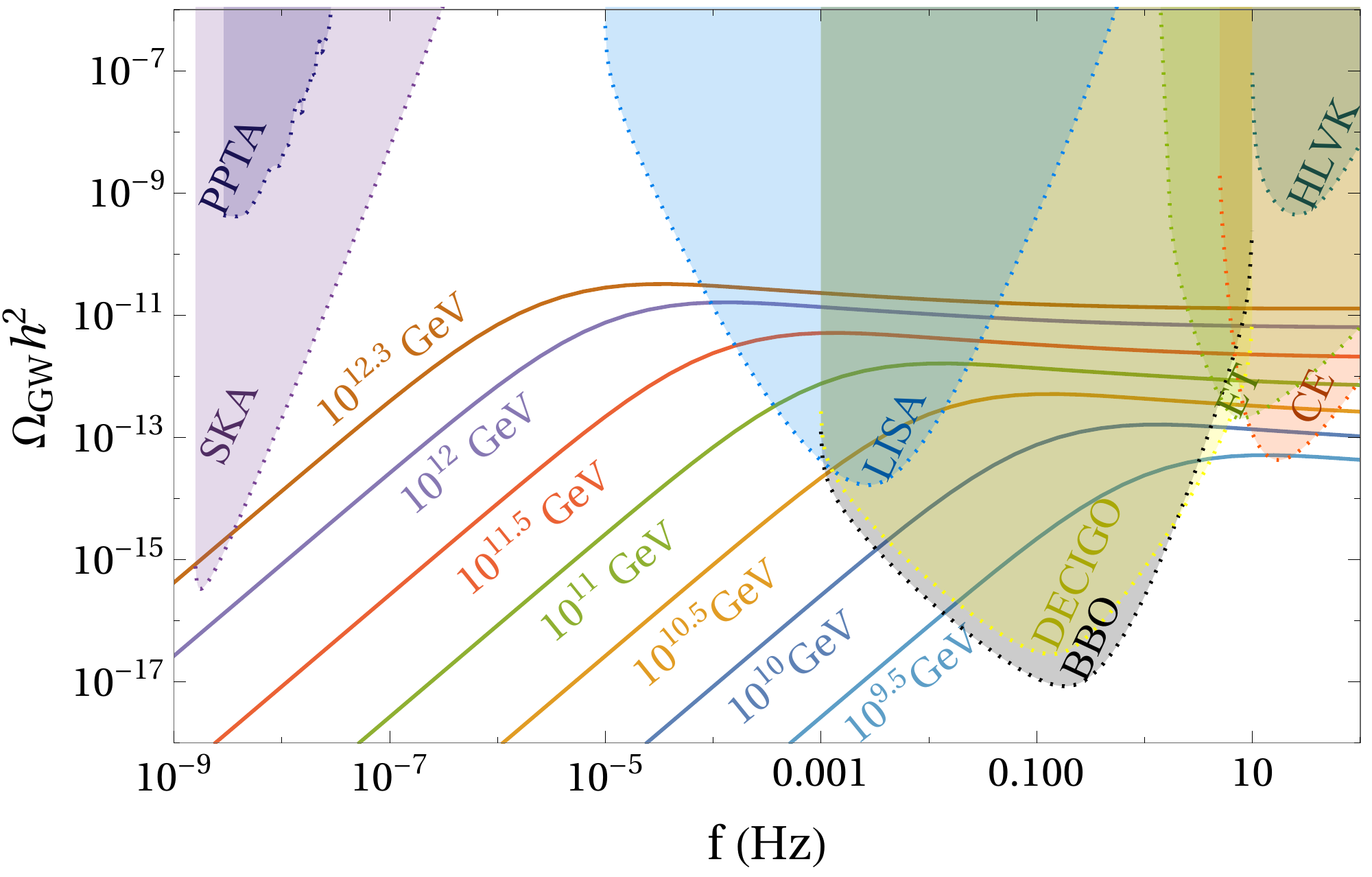}
\end{center}
\caption{Gravitational wave spectra from cosmic strings generated during the symmetry breaking at $\log_{10}(M_{II}/{\rm GeV})=[9.5,12.3]$. The sensitivity curves \cite{Thrane:2013oya, Schmitz:2020syl} for PPTA \cite{Shannon:2015ect} and various proposed experiments, namely, SKA \cite{5136190, Janssen:2014dka}, CE \cite{PhysRevLett.118.151105}, ET \cite{Mentasti:2020yyd}, LISA \cite{Bartolo:2016ami, amaroseoane2017laser}, DECIGO \cite{Sato_2017}, BBO \cite{Crowder:2005nr, Corbin:2005ny}, HLVK \cite{KAGRA:2013rdx}, etc. are shown on the plot.}\label{fig:GWs}
\end{figure}
%%%%%%%%%%%%%%%%%%%%%%%%%%%%%%%%%%%%%%%%%%%%%%%%%%%%%%%%%%%%%%%%%

The gravitational wave spectra for the breaking scales $M_{II}\in [10^{9.5},10^{12.3}]$ GeV are shown in Fig.~\ref{fig:GWs}. They satisfy the present PPTA bound \cite{Shannon:2015ect} and can be probed in various proposed experiments, including SKA \cite{5136190, Janssen:2014dka}, CE \cite{PhysRevLett.118.151105}, ET \cite{Mentasti:2020yyd}, LISA \cite{Bartolo:2016ami, amaroseoane2017laser}, DECIGO \cite{Sato_2017} and BBO \cite{Crowder:2005nr, Corbin:2005ny}. We have assumed, without loss of any generality, that the network of the string loops is present in the horizon from a very early time $t_F = 10^{-25}$ sec. In fact, in an inflationary universe driven by the Coleman-Weinberg potential of a real GUT-singlet \cite{Shafi:1983bd,Lazarides:1984pq}, inflation ends at a cosmic time $8.3\times 10^{-37}$ sec and the phase transitions occur during inflation only if the corresponding symmetry breaking scales $\gtrsim 10^{13}$ GeV \cite{Lazarides:2021uxv}. Therefore, the strings in the present case are produced after the end of inflation during the inflaton oscillation \cite{Chakrabortty:2020otp}. Needless to say, the new radiation temperature dominates over the Hawking temperature from the inflaton oscillations soon after inflation. Consequently, the Ginzburg criterion \cite{ginzburg} for a phase transition is governed by the radiation temperature which approaches the reheat temperature (see Ref.~\cite{Chakrabortty:2020otp}) at the reheat time $t_r\simeq 2.3 \times 10^{-25}$ sec.  The smaller loops formed during the inflaton oscillation era do not contribute to the gravitational wave background within the frequency range of nHz to kHz. Therefore, we can safely take $t_F = 10^{-25}$ sec to compute the GW spectra. 

%%%%%%%%%%%%%%%%%%%%%%%%%%%%%%%%%%%%%%%%%%%%%%%%%%%%%%%%%%%%%%%%%%%%
\section{Summary}\label{sec:summary}
We have discussed how the recent measurement of $m_W$ by CDF can be readily incorporated into a well-motivated axion model based on $SO(10)$ grand unification. No ad-hoc additional symmetries are imposed, in line with the spirit of the Standard Model. The axion symmetry breaking scalar field contains an $SU(2)_L$ triplet component that acquires a non-zero VEV through its mixing with the SM doublet. We show how the unification of the SM gauge couplings is preserved with an appropriate symmetry-breaking pattern of $SO(10)$. The proton lifetime is estimated to lie within the reach of future experiments.
The model contains two $10$-plets of fermions that are introduced to resolve the axion domain wall problem. An unbroken $Z_2$ gauge symmetry from $SO(10)$ ensures the presence of a stable intermediate-mass fermion from these $10$-plets which, in addition to the axion, is a plausible dark matter candidate.
The $Z_2$ symmetry also yields topologically stable intermediate scale cosmic strings whose gravitational wave spectrum we have also provided.
%%%%%%%%%%%%%%%%%%%%%%%%%%%%%%%%%%%%%%%%%%%%%%%%%%%%%%%%%%%%%%%%%%%%

%%%%%%%%%%%%%%%%
\section*{Acknowledgements}
%%%%%%%%%%%%%%%%
This work is supported by the Hellenic Foundation for Research and Innovation (H.F.R.I.) under the “First Call for H.F.R.I. Research Projects to support Faculty Members and Researchers and the procurement of high-cost research equipment grant” (Project Number: 2251). R.R. also acknowledges the National Research Foundation of Korea (NRF) grant funded by the Korean government (NRF-2020R1C1C1012452).

\bibliographystyle{mystyle}
\bibliography{GUT_TD}

\providecommand{\href}[2]{#2}\begingroup\raggedright\begin{thebibliography}{100}

\bibitem{Lazarides:2021los}
G.~Lazarides and Q.~Shafi, \emph{{Electroweak monopoles and magnetic dumbbells
  in grand unified theories}},
  \href{https://doi.org/10.1103/PhysRevD.103.095021}{\emph{Phys. Rev. D}
  {\bfseries 103} (2021) 095021}
  [\href{https://arxiv.org/abs/2102.07124}{{\ttfamily 2102.07124}}].

\bibitem{CDF:2022hxs}
{\scshape CDF} collaboration, \emph{{High-precision measurement of the W boson
  mass with the CDF II detector}},
  \href{https://doi.org/10.1126/science.abk1781}{\emph{Science} {\bfseries 376}
  (2022) 170}.

\bibitem{Zyla:2020zbs}
{\scshape Particle Data Group} collaboration, \emph{{Review of Particle
  Physics}}, \href{https://doi.org/10.1093/ptep/ptaa104}{\emph{PTEP} {\bfseries
  2020} (2020) 083C01}.

\bibitem{Lu:2022bgw}
C.-T.~Lu, L.~Wu, Y.~Wu and B.~Zhu, \emph{{Electroweak Precision Fit and New
  Physics in light of $W$ Boson Mass}},
  \href{https://arxiv.org/abs/2204.03796}{{\ttfamily 2204.03796}}.

\bibitem{deBlas:2022hdk}
J.~de~Blas, M.~Pierini, L.~Reina and L.~Silvestrini, \emph{{Impact of the
  recent measurements of the top-quark and W-boson masses on electroweak
  precision fits}},  \href{https://arxiv.org/abs/2204.04204}{{\ttfamily
  2204.04204}}.

\bibitem{Fan:2022dck}
Y.-Z.~Fan, T.-P.~Tang, Y.-L.S.~Tsai and L.~Wu, \emph{{Inert Higgs Dark Matter
  for New CDF W-boson Mass and Detection Prospects}},
  \href{https://arxiv.org/abs/2204.03693}{{\ttfamily 2204.03693}}.

\bibitem{Liu:2022jdq}
X.~Liu, S.-Y.~Guo, B.~Zhu and Y.~Li, \emph{{Unifying gravitational waves with
  $W$ boson, FIMP dark matter, and Majorana Seesaw mechanism}},
  \href{https://arxiv.org/abs/2204.04834}{{\ttfamily 2204.04834}}.

\bibitem{Athron:2022isz}
P.~Athron, M.~Bach, D.H.J.~Jacob, W.~Kotlarski, D.~St\"ockinger and A.~Voigt,
  \emph{{Precise calculation of the W boson pole mass beyond the Standard Model
  with FlexibleSUSY}},  \href{https://arxiv.org/abs/2204.05285}{{\ttfamily
  2204.05285}}.

\bibitem{Song:2022xts}
H.~Song, W.~Su and M.~Zhang, \emph{{Electroweak Phase Transition in 2HDM under
  Higgs, Z-pole, and W precision measurements}},
  \href{https://arxiv.org/abs/2204.05085}{{\ttfamily 2204.05085}}.

\bibitem{Cheung:2022zsb}
K.~Cheung, W.-Y.~Keung and P.-Y.~Tseng, \emph{{Iso-doublet Vector Leptoquark
  solution to the Muon $g-2$, $R_{K, K^*}$, $R_{D,D^*}$, and $W$-mass
  Anomalies}},  \href{https://arxiv.org/abs/2204.05942}{{\ttfamily
  2204.05942}}.

\bibitem{Endo:2022kiw}
M.~Endo and S.~Mishima, \emph{{New physics interpretation of $W$-boson mass
  anomaly}},  \href{https://arxiv.org/abs/2204.05965}{{\ttfamily 2204.05965}}.

\bibitem{Han:2022juu}
X.-F.~Han, F.~Wang, L.~Wang, J.M.~Yang and Y.~Zhang, \emph{{A joint explanation
  of W-mass and muon g-2 in 2HDM}},
  \href{https://arxiv.org/abs/2204.06505}{{\ttfamily 2204.06505}}.

\bibitem{Ahn:2022xeq}
Y.H.~Ahn, S.K.~Kang and R.~Ramos, \emph{{Implications of New CDF-II $W$ Boson
  Mass on Two Higgs Doublet Model}},
  \href{https://arxiv.org/abs/2204.06485}{{\ttfamily 2204.06485}}.

\bibitem{Perez:2022uil}
P.~Fileviez~Perez, H.H.~Patel and A.D.~Plascencia, \emph{{On the $W$-mass and
  New Higgs Bosons}},  \href{https://arxiv.org/abs/2204.07144}{{\ttfamily
  2204.07144}}.

\bibitem{Kawamura:2022uft}
J.~Kawamura, S.~Okawa and Y.~Omura, \emph{{$W$ boson mass and muon $g-2$ in a
  lepton portal dark matter model}},
  \href{https://arxiv.org/abs/2204.07022}{{\ttfamily 2204.07022}}.

\bibitem{Kanemura:2022ahw}
S.~Kanemura and K.~Yagyu, \emph{{Implication of the $W$ boson mass anomaly at
  CDF II in the Higgs triplet model with a mass difference}},
  \href{https://arxiv.org/abs/2204.07511}{{\ttfamily 2204.07511}}.

\bibitem{Nagao:2022oin}
K.I.~Nagao, T.~Nomura and H.~Okada, \emph{{A model explaining the new CDF II W
  boson mass linking to muon $g-2$ and dark matter}},
  \href{https://arxiv.org/abs/2204.07411}{{\ttfamily 2204.07411}}.

\bibitem{Mondal:2022xdy}
P.~Mondal, \emph{{Enhancement of the $\textbf{W}$ boson mass in the
  Georgi-Machacek model}},  \href{https://arxiv.org/abs/2204.07844}{{\ttfamily
  2204.07844}}.

\bibitem{Zhang:2022nnh}
K.-Y.~Zhang and W.-Z.~Feng, \emph{{Explaining $W$ boson mass anomaly and dark
  matter with a $U(1)$ dark sector}},
  \href{https://arxiv.org/abs/2204.08067}{{\ttfamily 2204.08067}}.

\bibitem{Carpenter:2022oyg}
L.M.~Carpenter, T.~Murphy and M.J.~Smylie, \emph{{Changing patterns in
  electroweak precision with new color-charged states: Oblique corrections and
  the $W$ boson mass}},  \href{https://arxiv.org/abs/2204.08546}{{\ttfamily
  2204.08546}}.

\bibitem{Popov:2022ldh}
O.~Popov and R.~Srivastava, \emph{{The Triplet Dirac Seesaw in the View of the
  Recent CDF-II W Mass Anomaly}},
  \href{https://arxiv.org/abs/2204.08568}{{\ttfamily 2204.08568}}.

\bibitem{Arcadi:2022dmt}
G.~Arcadi and A.~Djouadi, \emph{{The 2HD+a model for a combined explanation of
  the possible excesses in the CDF $\mathbf{M_W}$ measurement and
  $\mathbf{(g-2)_\mu}$ with Dark Matter}},
  \href{https://arxiv.org/abs/2204.08406}{{\ttfamily 2204.08406}}.

\bibitem{Chowdhury:2022moc}
T.A.~Chowdhury, J.~Heeck, S.~Saad and A.~Thapa, \emph{{$W$ boson mass shift and
  muon magnetic moment in the Zee model}},
  \href{https://arxiv.org/abs/2204.08390}{{\ttfamily 2204.08390}}.

\bibitem{Borah:2022obi}
D.~Borah, S.~Mahapatra, D.~Nanda and N.~Sahu, \emph{{Type II Dirac Seesaw with
  Observable $\Delta N_{eff}$ in the light of W-mass Anomaly}},
  \href{https://arxiv.org/abs/2204.08266}{{\ttfamily 2204.08266}}.

\bibitem{Du:2022fqv}
M.~Du, Z.~Liu and P.~Nath, \emph{{CDF W mass anomaly from a dark sector with a
  Stueckelberg-Higgs portal}},
  \href{https://arxiv.org/abs/2204.09024}{{\ttfamily 2204.09024}}.

\bibitem{Ghorbani:2022vtv}
K.~Ghorbani and P.~Ghorbani, \emph{{$W$-Boson Mass Anomaly from Scale Invariant
  2HDM}},  \href{https://arxiv.org/abs/2204.09001}{{\ttfamily 2204.09001}}.

\bibitem{Asadi:2022xiy}
P.~Asadi, C.~Cesarotti, K.~Fraser, S.~Homiller and A.~Parikh, \emph{{Oblique
  Lessons from the $W$ Mass Measurement at CDF II}},
  \href{https://arxiv.org/abs/2204.05283}{{\ttfamily 2204.05283}}.

\bibitem{Bhaskar:2022vgk}
A.~Bhaskar, A.A.~Madathil, T.~Mandal and S.~Mitra, \emph{{Combined explanation
  of $W$-mass, muon $g-2$, $R_{K^{(*)}}$ and $R_{D^{(*)}}$ anomalies in a
  singlet-triplet scalar leptoquark model}},
  \href{https://arxiv.org/abs/2204.09031}{{\ttfamily 2204.09031}}.

\bibitem{Babu:2022pdn}
K.S.~Babu, S.~Jana and V.P.~K., \emph{{Correlating $W$-Boson Mass Shift with
  Muon ${g-2}$ in the 2HDM}},
  \href{https://arxiv.org/abs/2204.05303}{{\ttfamily 2204.05303}}.

\bibitem{Ghoshal:2022vzo}
A.~Ghoshal, N.~Okada, S.~Okada, D.~Raut, Q.~Shafi and A.~Thapa, \emph{{Type III
  seesaw with R-parity violation in light of $m_W$ (CDF)}},
  \href{https://arxiv.org/abs/2204.07138}{{\ttfamily 2204.07138}}.

\bibitem{Arias-Aragon:2022ats}
F.~Arias-Arag\'on, E.~Fern\'andez-Mart\'\i{}nez, M.~Gonz\'alez-L\'opez and
  L.~Merlo, \emph{{Dynamical Minimal Flavour Violating Inverse Seesaw}},
  \href{https://arxiv.org/abs/2204.04672}{{\ttfamily 2204.04672}}.

\bibitem{Sakurai:2022hwh}
K.~Sakurai, F.~Takahashi and W.~Yin, \emph{{Singlet extensions and W boson mass
  in the light of the CDF II result}},
  \href{https://arxiv.org/abs/2204.04770}{{\ttfamily 2204.04770}}.

\bibitem{Gu:2022htv}
J.~Gu, Z.~Liu, T.~Ma and J.~Shu, \emph{{Speculations on the W-Mass Measurement
  at CDF}},  \href{https://arxiv.org/abs/2204.05296}{{\ttfamily 2204.05296}}.

\bibitem{Batra:2022org}
A.~Batra, S.~K.~A., S.~Mandal and R.~Srivastava, \emph{{W boson mass in
  Singlet-Triplet Scotogenic dark matter model}},
  \href{https://arxiv.org/abs/2204.09376}{{\ttfamily 2204.09376}}.

\bibitem{Baek:2022agi}
S.~Baek, \emph{{Implications of CDF $W$-mass and $(g-2)_\mu$ on
  $U(1)_{L_\mu-L_\tau}$ model}},
  \href{https://arxiv.org/abs/2204.09585}{{\ttfamily 2204.09585}}.

\bibitem{Borah:2022zim}
D.~Borah, S.~Mahapatra and N.~Sahu, \emph{{Singlet-Doublet Fermion Origin of
  Dark Matter, Neutrino Mass and W-Mass Anomaly}},
  \href{https://arxiv.org/abs/2204.09671}{{\ttfamily 2204.09671}}.

\bibitem{Heeck:2022fvl}
J.~Heeck, \emph{{W-boson mass in the triplet seesaw model}},
  \href{https://arxiv.org/abs/2204.10274}{{\ttfamily 2204.10274}}.

\bibitem{Addazi:2022fbj}
A.~Addazi, A.~Marciano, A.P.~Morais, R.~Pasechnik and H.~Yang, \emph{{CDF II
  $W$-mass anomaly faces first-order electroweak phase transition}},
  \href{https://arxiv.org/abs/2204.10315}{{\ttfamily 2204.10315}}.

\bibitem{Cheng:2022aau}
Y.~Cheng, X.-G.~He, F.~Huang, J.~Sun and Z.-P.~Xing, \emph{{Dark photon kinetic
  mixing effects for CDF W mass excess}},
  \href{https://arxiv.org/abs/2204.10156}{{\ttfamily 2204.10156}}.

\bibitem{Crivellin:2022fdf}
A.~Crivellin, M.~Kirk, T.~Kitahara and F.~Mescia, \emph{{Correlating $t\to cZ$
  to the $W$ Mass and $B$ Physics with Vector-Like Quarks}},
  \href{https://arxiv.org/abs/2204.05962}{{\ttfamily 2204.05962}}.

\bibitem{Lee:2022gyf}
S.~Lee, K.~Cheung, J.~Kim, C.-T.~Lu and J.~Song, \emph{{Status of the
  two-Higgs-doublet model in light of the CDF $m_W$ measurement}},
  \href{https://arxiv.org/abs/2204.10338}{{\ttfamily 2204.10338}}.

\bibitem{Batra:2022pej}
A.~Batra, S.K.~A, S.~Mandal, H.~Prajapati and R.~Srivastava, \emph{{CDF-II $W$
  Boson Mass Anomaly in the Canonical Scotogenic Neutrino-Dark Matter Model}},
  \href{https://arxiv.org/abs/2204.11945}{{\ttfamily 2204.11945}}.

\bibitem{Benbrik:2022dja}
R.~Benbrik, M.~Boukidi and B.~Manaut, \emph{{$W$-mass and 96 GeV excess in
  type-III 2HDM}},  \href{https://arxiv.org/abs/2204.11755}{{\ttfamily
  2204.11755}}.

\bibitem{Cai:2022cti}
C.~Cai, D.~Qiu, Y.-L.~Tang, Z.-H.~Yu and H.-H.~Zhang, \emph{{Corrections to
  electroweak precision observables from mixings of an exotic vector boson in
  light of the CDF $W$-mass anomaly}},
  \href{https://arxiv.org/abs/2204.11570}{{\ttfamily 2204.11570}}.

\bibitem{Zhou:2022cql}
Q.~Zhou and X.-F.~Han, \emph{{The CDF W-mass, muon g-2, and dark matter in a
  $U(1)_{L_\mu-L_\tau}$ model with vector-like leptons}},
  \href{https://arxiv.org/abs/2204.13027}{{\ttfamily 2204.13027}}.

\bibitem{Zhu:2022tpr}
C.-R.~Zhu, M.-Y.~Cui, Z.-Q.~Xia, Z.-H.~Yu, X.~Huang, Q.~Yuan et~al., \emph{{GeV
  antiproton/gamma-ray excesses and the $W$-boson mass anomaly: three faces of
  $\sim 60-70$ GeV dark matter particle?}},
  \href{https://arxiv.org/abs/2204.03767}{{\ttfamily 2204.03767}}.

\bibitem{Wang:2022dte}
J.-W.~Wang, X.-J.~Bi, P.-F.~Yin and Z.-H.~Yu, \emph{{Electroweak dark matter
  model accounting for the CDF $W$-mass anomaly}},
  \href{https://arxiv.org/abs/2205.00783}{{\ttfamily 2205.00783}}.

\bibitem{Dcruz:2022dao}
R.~Dcruz and A.~Thapa, \emph{{$W$ boson mass, dark matter and $(g-2)_\ell$ in
  ScotoZee neutrino mass model}},
  \href{https://arxiv.org/abs/2205.02217}{{\ttfamily 2205.02217}}.

\bibitem{Li:2022gwc}
X.-Q.~Li, Z.-J.~Xie, Y.-D.~Yang and X.-B.~Yuan, \emph{{Correlating the CDF
  $W$-boson mass shift with the $b \to s \ell^+ \ell^-$ anomalies}},
  \href{https://arxiv.org/abs/2205.02205}{{\ttfamily 2205.02205}}.

\bibitem{He:2022zjz}
S.-P.~He, \emph{{A leptoquark and vector-like quark extended model for the
  simultaneous explanation of the $W$ boson mass and muon $g-2$ anomalies}},
  \href{https://arxiv.org/abs/2205.02088}{{\ttfamily 2205.02088}}.

\bibitem{Kim:2022hvh}
J.~Kim, S.~Lee, P.~Sanyal and J.~Song, \emph{{CDF $m_W$ and the muon $g-2$
  through the Higgs-phobic light pseudoscalar in type-X two-Higgs-doublet
  model}},  \href{https://arxiv.org/abs/2205.01701}{{\ttfamily 2205.01701}}.

\bibitem{Evans:2022dgq}
J.L.~Evans, T.T.~Yanagida and N.~Yokozaki, \emph{{W boson mass anomaly and
  grand unification}},  \href{https://arxiv.org/abs/2205.03877}{{\ttfamily
  2205.03877}}.

\bibitem{Chowdhury:2022dps}
T.A.~Chowdhury and S.~Saad, \emph{{Leptoquark-vectorlike quark model for $m_W$
  (CDF), $(g-2)_\mu$, $R_{K^{(\ast)}}$ anomalies and neutrino mass}},
  \href{https://arxiv.org/abs/2205.03917}{{\ttfamily 2205.03917}}.

\bibitem{Kim:2022zhj}
S.-S.~Kim, H.M.~Lee, A.~Menkara and K.~Yamashita, \emph{{The $SU(2)_D$ lepton
  portals for muon $g-2$, $W$ boson mass and dark matter}},
  \href{https://arxiv.org/abs/2205.04016}{{\ttfamily 2205.04016}}.

\bibitem{Athron:2022qpo}
P.~Athron, A.~Fowlie, C.-T.~Lu, L.~Wu, Y.~Wu and B.~Zhu, \emph{{The $W$ boson
  Mass and Muon $g-2$: Hadronic Uncertainties or New Physics?}},
  \href{https://arxiv.org/abs/2204.03996}{{\ttfamily 2204.03996}}.

\bibitem{DiLuzio:2022xns}
L.~Di~Luzio, R.~Gr\"ober and P.~Paradisi, \emph{{Higgs physics confronts the
  $M_W$ anomaly}},  \href{https://arxiv.org/abs/2204.05284}{{\ttfamily
  2204.05284}}.

\bibitem{Heckman:2022the}
J.J.~Heckman, \emph{{Extra $W$-Boson Mass from a D3-Brane}},
  \href{https://arxiv.org/abs/2204.05302}{{\ttfamily 2204.05302}}.

\bibitem{Chen:2022ocr}
T.-K.~Chen, C.-W.~Chiang and K.~Yagyu, \emph{{Explanation of the $W$ mass shift
  at CDF II in the Georgi-Machacek Model}},
  \href{https://arxiv.org/abs/2204.12898}{{\ttfamily 2204.12898}}.

\bibitem{Kim:2022xuo}
J.~Kim, \emph{{Compatibility of muon $g-2$, $W$ mass anomaly in type-X 2HDM}},
  \href{https://arxiv.org/abs/2205.01437}{{\ttfamily 2205.01437}}.

\bibitem{Barman:2022qix}
B.~Barman, A.~Das and S.~Sengupta, \emph{{New $W$-Boson mass in the light of
  doubly warped braneworld model}},
  \href{https://arxiv.org/abs/2205.01699}{{\ttfamily 2205.01699}}.

\bibitem{Holman:1982tb}
R.~Holman, G.~Lazarides and Q.~Shafi, \emph{{Axions and the Dark Matter of the
  Universe}}, \href{https://doi.org/10.1103/PhysRevD.27.995}{\emph{Phys. Rev.
  D} {\bfseries 27} (1983) 995}.

\bibitem{Lazarides:2020frf}
G.~Lazarides and Q.~Shafi, \emph{{Axion Model with Intermediate Scale Fermionic
  Dark Matter}},
  \href{https://doi.org/10.1016/j.physletb.2020.135603}{\emph{Phys. Lett. B}
  {\bfseries 807} (2020) 135603}
  [\href{https://arxiv.org/abs/2004.11560}{{\ttfamily 2004.11560}}].

\bibitem{Babu:2015bna}
K.S.~Babu and S.~Khan, \emph{{Minimal nonsupersymmetric $SO(10)$ model: Gauge
  coupling unification, proton decay, and fermion masses}},
  \href{https://doi.org/10.1103/PhysRevD.92.075018}{\emph{Phys. Rev. D}
  {\bfseries 92} (2015) 075018}
  [\href{https://arxiv.org/abs/1507.06712}{{\ttfamily 1507.06712}}].

\bibitem{Giddings:1987cg}
S.B.~Giddings and A.~Strominger, \emph{{Axion Induced Topology Change in
  Quantum Gravity and String Theory}},
  \href{https://doi.org/10.1016/0550-3213(88)90446-4}{\emph{Nucl. Phys. B}
  {\bfseries 306} (1988) 890}.

\bibitem{Lee:1988ge}
K.-M.~Lee, \emph{{Wormholes and Goldstone Bosons}},
  \href{https://doi.org/10.1103/PhysRevLett.61.263}{\emph{Phys. Rev. Lett.}
  {\bfseries 61} (1988) 263}.

\bibitem{Alvey:2020nyh}
J.~Alvey and M.~Escudero, \emph{{The axion quality problem: global symmetry
  breaking and wormholes}},
  \href{https://doi.org/10.1007/JHEP01(2021)032}{\emph{JHEP} {\bfseries 01}
  (2021) 032} [\href{https://arxiv.org/abs/2009.03917}{{\ttfamily
  2009.03917}}].

\bibitem{Lazarides:1985bj}
G.~Lazarides, C.~Panagiotakopoulos and Q.~Shafi, \emph{{Phenomenology and
  Cosmology With Superstrings}},
  \href{https://doi.org/10.1103/PhysRevLett.56.432}{\emph{Phys. Rev. Lett.}
  {\bfseries 56} (1986) 432}.

\bibitem{DiLuzio:2020qio}
L.~Di~Luzio, \emph{{Accidental SO(10) axion from gauged flavour}},
  \href{https://doi.org/10.1007/JHEP11(2020)074}{\emph{JHEP} {\bfseries 11}
  (2020) 074} [\href{https://arxiv.org/abs/2008.09119}{{\ttfamily
  2008.09119}}].

\bibitem{Planck:2018vyg}
{\scshape Planck} collaboration, \emph{{Planck 2018 results. VI. Cosmological
  parameters}},
  \href{https://doi.org/10.1051/0004-6361/201833910}{\emph{Astron. Astrophys.}
  {\bfseries 641} (2020) A6}
  [\href{https://arxiv.org/abs/1807.06209}{{\ttfamily 1807.06209}}].

\bibitem{Senoguz:2015lba}
V.N.~\c{S}eno\u{g}uz and Q.~Shafi, \emph{{Primordial monopoles, proton decay,
  gravity waves and GUT inflation}},
  \href{https://doi.org/10.1016/j.physletb.2015.11.037}{\emph{Phys. Lett. B}
  {\bfseries 752} (2016) 169}
  [\href{https://arxiv.org/abs/1510.04442}{{\ttfamily 1510.04442}}].

\bibitem{Chakrabortty:2020otp}
J.~Chakrabortty, G.~Lazarides, R.~Maji and Q.~Shafi, \emph{{Primordial
  Monopoles and Strings, Inflation, and Gravity Waves}},
  \href{https://doi.org/10.1007/JHEP02(2021)114}{\emph{JHEP} {\bfseries 02}
  (2021) 114} [\href{https://arxiv.org/abs/2011.01838}{{\ttfamily
  2011.01838}}].

\bibitem{Jones:1981we}
D.R.T.~Jones, \emph{{The Two Loop $\beta$ Function for a $G_1 \times G_2$ Gauge
  Theory}}, \href{https://doi.org/10.1103/PhysRevD.25.581}{\emph{Phys. Rev. D}
  {\bfseries 25} (1982) 581}.

\bibitem{Chakrabortty:2017mgi}
J.~Chakrabortty, R.~Maji, S.K.~Patra, T.~Srivastava and S.~Mohanty,
  \emph{{Roadmap of left-right models based on GUTs}},
  \href{https://doi.org/10.1103/PhysRevD.97.095010}{\emph{Phys. Rev. D}
  {\bfseries 97} (2018) 095010}
  [\href{https://arxiv.org/abs/1711.11391}{{\ttfamily 1711.11391}}].

\bibitem{Chakrabortty:2019fov}
J.~Chakrabortty, R.~Maji and S.F.~King, \emph{{Unification, Proton Decay and
  Topological Defects in non-SUSY GUTs with Thresholds}},
  \href{https://doi.org/10.1103/PhysRevD.99.095008}{\emph{Phys. Rev. D}
  {\bfseries 99} (2019) 095008}
  [\href{https://arxiv.org/abs/1901.05867}{{\ttfamily 1901.05867}}].

\bibitem{Super-Kamiokande:2020wjk}
{\scshape Super-Kamiokande} collaboration, \emph{{Search for proton decay via
  $p\to e^+\pi^0$ and $p\to \mu^+\pi^0$ with an enlarged fiducial volume in
  Super-Kamiokande I-IV}},
  \href{https://doi.org/10.1103/PhysRevD.102.112011}{\emph{Phys. Rev. D}
  {\bfseries 102} (2020) 112011}
  [\href{https://arxiv.org/abs/2010.16098}{{\ttfamily 2010.16098}}].

\bibitem{Dealtry:2019ldr}
{\scshape Hyper-Kamiokande} collaboration, \emph{{Hyper-Kamiokande}},  in
  \emph{{Prospects in Neutrino Physics}}, 4, 2019
  [\href{https://arxiv.org/abs/1904.10206}{{\ttfamily 1904.10206}}].

\bibitem{Ambrosio:2002qq}
{\scshape MACRO} collaboration, \emph{{Final results of magnetic monopole
  searches with the MACRO experiment}},
  \href{https://doi.org/10.1140/epjc/s2002-01046-9}{\emph{Eur. Phys. J. C}
  {\bfseries 25} (2002) 511}
  [\href{https://arxiv.org/abs/hep-ex/0207020}{{\ttfamily hep-ex/0207020}}].

\bibitem{Preskill:1982cy}
J.~Preskill, M.B.~Wise and F.~Wilczek, \emph{{Cosmology of the Invisible
  Axion}}, \href{https://doi.org/10.1016/0370-2693(83)90637-8}{\emph{Phys.
  Lett. B} {\bfseries 120} (1983) 127}.

\bibitem{Abbott:1982af}
L.F.~Abbott and P.~Sikivie, \emph{{A Cosmological Bound on the Invisible
  Axion}}, \href{https://doi.org/10.1016/0370-2693(83)90638-X}{\emph{Phys.
  Lett. B} {\bfseries 120} (1983) 133}.

\bibitem{Stecker:1982ws}
F.W.~Stecker and Q.~Shafi, \emph{{The Evolution of Structure in the Universe
  From Axions}}, \href{https://doi.org/10.1103/PhysRevLett.50.928}{\emph{Phys.
  Rev. Lett.} {\bfseries 50} (1983) 928}.

\bibitem{Visinelli:2009zm}
L.~Visinelli and P.~Gondolo, \emph{{Dark Matter Axions Revisited}},
  \href{https://doi.org/10.1103/PhysRevD.80.035024}{\emph{Phys. Rev. D}
  {\bfseries 80} (2009) 035024}
  [\href{https://arxiv.org/abs/0903.4377}{{\ttfamily 0903.4377}}].

\bibitem{Hagmann:2000ja}
C.~Hagmann, S.~Chang and P.~Sikivie, \emph{{Axion radiation from strings}},
  \href{https://doi.org/10.1103/PhysRevD.63.125018}{\emph{Phys. Rev. D}
  {\bfseries 63} (2001) 125018}
  [\href{https://arxiv.org/abs/hep-ph/0012361}{{\ttfamily hep-ph/0012361}}].

\bibitem{Dimopoulos:2003ii}
K.~Dimopoulos, G.~Lazarides, D.~Lyth and R.~Ruiz~de Austri, \emph{{The
  Peccei-Quinn field as curvaton}},
  \href{https://doi.org/10.1088/1126-6708/2003/05/057}{\emph{JHEP} {\bfseries
  05} (2003) 057} [\href{https://arxiv.org/abs/hep-ph/0303154}{{\ttfamily
  hep-ph/0303154}}].

\bibitem{Olmez:2010bi}
S.~Olmez, V.~Mandic and X.~Siemens, \emph{{Gravitational-Wave Stochastic
  Background from Kinks and Cusps on Cosmic Strings}},
  \href{https://doi.org/10.1103/PhysRevD.81.104028}{\emph{Phys. Rev. D}
  {\bfseries 81} (2010) 104028}
  [\href{https://arxiv.org/abs/1004.0890}{{\ttfamily 1004.0890}}].

\bibitem{Cui:2019kkd}
Y.~Cui, M.~Lewicki and D.E.~Morrissey, \emph{{Gravitational Wave Bursts as
  Harbingers of Cosmic Strings Diluted by Inflation}},
  \href{https://doi.org/10.1103/PhysRevLett.125.211302}{\emph{Phys. Rev. Lett.}
  {\bfseries 125} (2020) 211302}
  [\href{https://arxiv.org/abs/1912.08832}{{\ttfamily 1912.08832}}].

\bibitem{Auclair:2019wcv}
P.~Auclair et~al., \emph{{Probing the gravitational wave background from cosmic
  strings with LISA}},
  \href{https://doi.org/10.1088/1475-7516/2020/04/034}{\emph{JCAP} {\bfseries
  04} (2020) 034} [\href{https://arxiv.org/abs/1909.00819}{{\ttfamily
  1909.00819}}].

\bibitem{Blanco-Pillado:2013qja}
J.J.~Blanco-Pillado, K.D.~Olum and B.~Shlaer, \emph{{The number of cosmic
  string loops}}, \href{https://doi.org/10.1103/PhysRevD.89.023512}{\emph{Phys.
  Rev. D} {\bfseries 89} (2014) 023512}
  [\href{https://arxiv.org/abs/1309.6637}{{\ttfamily 1309.6637}}].

\bibitem{Blanco-Pillado:2017oxo}
J.J.~Blanco-Pillado and K.D.~Olum, \emph{{Stochastic gravitational wave
  background from smoothed cosmic string loops}},
  \href{https://doi.org/10.1103/PhysRevD.96.104046}{\emph{Phys. Rev. D}
  {\bfseries 96} (2017) 104046}
  [\href{https://arxiv.org/abs/1709.02693}{{\ttfamily 1709.02693}}].

\bibitem{LIGOScientific:2021nrg}
{\scshape LIGO Scientific, Virgo, KAGRA} collaboration, \emph{{Constraints on
  Cosmic Strings Using Data from the Third Advanced LIGO\textendash{}Virgo
  Observing Run}},
  \href{https://doi.org/10.1103/PhysRevLett.126.241102}{\emph{Phys. Rev. Lett.}
  {\bfseries 126} (2021) 241102}
  [\href{https://arxiv.org/abs/2101.12248}{{\ttfamily 2101.12248}}].

\bibitem{Thrane:2013oya}
E.~Thrane and J.D.~Romano, \emph{{Sensitivity curves for searches for
  gravitational-wave backgrounds}},
  \href{https://doi.org/10.1103/PhysRevD.88.124032}{\emph{Phys. Rev. D}
  {\bfseries 88} (2013) 124032}
  [\href{https://arxiv.org/abs/1310.5300}{{\ttfamily 1310.5300}}].

\bibitem{Schmitz:2020syl}
K.~Schmitz, \emph{{New Sensitivity Curves for Gravitational-Wave Signals from
  Cosmological Phase Transitions}},
  \href{https://doi.org/10.1007/JHEP01(2021)097}{\emph{JHEP} {\bfseries 01}
  (2021) 097} [\href{https://arxiv.org/abs/2002.04615}{{\ttfamily
  2002.04615}}].

\bibitem{Shannon:2015ect}
R.~Shannon et~al., \emph{{Gravitational waves from binary supermassive black
  holes missing in pulsar observations}},
  \href{https://doi.org/10.1126/science.aab1910}{\emph{Science} {\bfseries 349}
  (2015) 1522} [\href{https://arxiv.org/abs/1509.07320}{{\ttfamily
  1509.07320}}].

\bibitem{5136190}
P.E.~{Dewdney}, P.J.~{Hall}, R.T.~{Schilizzi} and T.J.L.W.~{Lazio}, \emph{The
  square kilometre array},
  \href{https://doi.org/10.1109/JPROC.2009.2021005}{\emph{Proceedings of the
  IEEE} {\bfseries 97} (2009) 1482}.

\bibitem{Janssen:2014dka}
G.~Janssen et~al., \emph{{Gravitational wave astronomy with the SKA}},
  \href{https://doi.org/10.22323/1.215.0037}{\emph{PoS} {\bfseries AASKA14}
  (2015) 037} [\href{https://arxiv.org/abs/1501.00127}{{\ttfamily
  1501.00127}}].

\bibitem{PhysRevLett.118.151105}
T.~Regimbau, M.~Evans, N.~Christensen, E.~Katsavounidis, B.~Sathyaprakash and
  S.~Vitale, \emph{Digging deeper: Observing primordial gravitational waves
  below the binary-black-hole-produced stochastic background},
  \href{https://doi.org/10.1103/PhysRevLett.118.151105}{\emph{Phys. Rev. Lett.}
  {\bfseries 118} (2017) 151105}.

\bibitem{Mentasti:2020yyd}
G.~Mentasti and M.~Peloso, \emph{{ET sensitivity to the anisotropic Stochastic
  Gravitational Wave Background}},
  \href{https://doi.org/10.1088/1475-7516/2021/03/080}{\emph{JCAP} {\bfseries
  03} (2021) 080} [\href{https://arxiv.org/abs/2010.00486}{{\ttfamily
  2010.00486}}].

\bibitem{Bartolo:2016ami}
N.~Bartolo et~al., \emph{{Science with the space-based interferometer LISA. IV:
  Probing inflation with gravitational waves}},
  \href{https://doi.org/10.1088/1475-7516/2016/12/026}{\emph{JCAP} {\bfseries
  12} (2016) 026} [\href{https://arxiv.org/abs/1610.06481}{{\ttfamily
  1610.06481}}].

\bibitem{amaroseoane2017laser}
P.~Amaro-Seoane et~al., \emph{Laser interferometer space antenna},
  \href{https://arxiv.org/abs/1702.00786}{{\ttfamily 1702.00786}}.

\bibitem{Sato_2017}
S.~Sato et~al., \emph{The status of {DECIGO}},
  \href{https://doi.org/10.1088/1742-6596/840/1/012010}{\emph{Journal of
  Physics: Conference Series} {\bfseries 840} (2017) 012010}.

\bibitem{Crowder:2005nr}
J.~Crowder and N.J.~Cornish, \emph{{Beyond LISA: Exploring future gravitational
  wave missions}},
  \href{https://doi.org/10.1103/PhysRevD.72.083005}{\emph{Phys. Rev. D}
  {\bfseries 72} (2005) 083005}
  [\href{https://arxiv.org/abs/gr-qc/0506015}{{\ttfamily gr-qc/0506015}}].

\bibitem{Corbin:2005ny}
V.~Corbin and N.J.~Cornish, \emph{{Detecting the cosmic gravitational wave
  background with the big bang observer}},
  \href{https://doi.org/10.1088/0264-9381/23/7/014}{\emph{Class. Quant. Grav.}
  {\bfseries 23} (2006) 2435}
  [\href{https://arxiv.org/abs/gr-qc/0512039}{{\ttfamily gr-qc/0512039}}].

\bibitem{KAGRA:2013rdx}
{\scshape KAGRA, LIGO Scientific, Virgo, VIRGO} collaboration, \emph{{Prospects
  for observing and localizing gravitational-wave transients with Advanced
  LIGO, Advanced Virgo and KAGRA}},
  \href{https://doi.org/10.1007/s41114-020-00026-9}{\emph{Living Rev. Rel.}
  {\bfseries 21} (2018) 3} [\href{https://arxiv.org/abs/1304.0670}{{\ttfamily
  1304.0670}}].

\bibitem{Shafi:1983bd}
Q.~Shafi and A.~Vilenkin, \emph{{Inflation with SU(5)}},
  \href{https://doi.org/10.1103/PhysRevLett.52.691}{\emph{Phys. Rev. Lett.}
  {\bfseries 52} (1984) 691}.

\bibitem{Lazarides:1984pq}
G.~Lazarides and Q.~Shafi, \emph{{Extended Structures at Intermediate Scales in
  an Inflationary Cosmology}},
  \href{https://doi.org/10.1016/0370-2693(84)91605-8}{\emph{Phys. Lett. B}
  {\bfseries 148} (1984) 35}.

\bibitem{Lazarides:2021uxv}
G.~Lazarides, R.~Maji and Q.~Shafi, \emph{{Cosmic strings, inflation, and
  gravity waves}},
  \href{https://doi.org/10.1103/PhysRevD.104.095004}{\emph{Phys. Rev. D}
  {\bfseries 104} (2021) 095004}
  [\href{https://arxiv.org/abs/2104.02016}{{\ttfamily 2104.02016}}].

\bibitem{ginzburg}
V.L.~Ginzburg, \emph{{Some Remarks on Phase Transitions of the Second Kind and
  the Microscopic theory of Ferroelectric Materials}}, {\emph{{Soviet Phys.
  Solid State}} {\bfseries {2}} ({1961}) 1824}.

\end{thebibliography}\endgroup
%%%%%%%%%%%%%%%%%%%%%%%%%%

\end{document}